\documentclass[prd,preprint,tightenlines,floatfix,
showpacs,preprintnumbers,nofootinbib,eqsecnum]{revtex4}

\usepackage[T1]{fontenc}		

\usepackage{amsmath,amsfonts,amssymb,amstext,mathrsfs}
\usepackage{mathpazo}

\usepackage[dvips]{graphicx}
\usepackage{epsf,float}
\usepackage{revsymb}

\usepackage{dcolumn}
\usepackage{braket}
\usepackage{color,xcolor}
\usepackage{graphicx}
\usepackage{subfigure}
\usepackage{multirow}
\usepackage{tabularx}
\usepackage{pstricks}
\usepackage[section]{placeins}
\usepackage{booktabs}
\usepackage{array}

\usepackage{hyperref}

\def\Pom{I\!\!P}
\def\Reg{I\!\!R}

\begin{document}

\title{Revision of absorption corrections for the $p p \to p p \pi^{+} \pi^{-}$ process}

\author{Piotr Lebiedowicz}
\email{Piotr.Lebiedowicz@ifj.edu.pl}
\author{Antoni Szczurek \footnote{Also at University of Rzesz\'ow, PL-35-959 Rzesz\'ow, Poland.}}
\email{Antoni.Szczurek@ifj.edu.pl}
\affiliation{Institute of Nuclear Physics PAN, PL-31-342 Krak\'ow, Poland}

\vspace{4cm}

\begin{abstract}
We include new additional absorption corrections into
the Lebiedowicz-Szczurek (non-resonant) model for
$p p \to p p \pi^+ \pi^-$ or $p \bar p \to p \bar p \pi^+ \pi^-$
processes. They are related to the $\pi N$ nonperturbative interaction 
in the final state of the reaction.
The role of the absorption corrections is quantified for several
differential distributions for $\sqrt{s}$ = 0.2, 1.96, 7, and 8 TeV.
The new absorption corrections lead to further decrease of 
the cross section by about a factor of two.
They change the shape of some distributions 
($d \sigma / dt$, $d \sigma / dp_{t,p}$, $d \sigma/d \phi_{pp}$) 
but leave almost unchanged shape of other distributions 
($d \sigma / dM_{\pi \pi}$, $d \sigma / dy_{\pi}$, $d \sigma /dp_{t,\pi}$, 
$d \sigma/ d \phi_{\pi \pi}$). 
The effect may have important impact on the interpretation
of the recent STAR and CDF data as well as the forthcoming data 
of the ALICE, ATLAS + ALFA and CMS + TOTEM collaborations.
\end{abstract}


\maketitle

\section{Introduction}

There is a growing experimental and theoretical interest 
in understanding of soft hadronic processes at high energy; 
for reviews see e.g.~\cite{Albrow:2010yb,Lebiedowicz:thesis,Albrow:2014bta}
and references therein.
One of the reaction which can be relatively easy to measure is 
$p p \to p \pi^+ \pi^- p$ or $p \bar p \to p \pi^+ \pi^- \bar{p}$ 
(four charged particles in the final state). 
There are recently several experimental projects by the COMPASS \cite{Austregesilo:2013yxa,Austregesilo:2014oxa}, 
STAR \cite{Adamczyk:2014ofa}, CDF \cite{Albrow_Project_new,Aaltonen:2015uva},
ALICE \cite{Schicker:2012nn,Schicker:2014aoa}, ATLAS \cite{Staszewski:2011bg},
and CMS \cite{Osterberg:2014mta} collaborations which will measure 
differential cross sections for the reaction(s).
Here we wish to compare predictions from the Lebiedowicz-Szczurek model 
with the recent STAR and CDF data.

The principal reason for studying central exclusive production of mesons 
is to search for glueballs \cite{Szczurek:2009yk}.
There is some evidence from an analysis of the decay modes of the scalar states observed, 
that the lightest scalar glueball manifests itself through 
the mixing with nearby $q \bar{q}$ states \cite{Ochs:2013gi,Kirk:2014nwa}.
The exclusive production of lower mass scalar and pseudoscalar resonances
within a tensor pomeron approach \cite{Ewerz:2013kda}
was recently examined in \cite{Lebiedowicz:2013ika}.
Resonant ($\rho^0 \to \pi^{+}\pi^{-}$) and non-resonant (Drell-S\"oding)
photon-pomeron/reggeon production was studied in \cite{Lebiedowicz:2014bea}.
In Refs.~\cite{Lebiedowicz:2011nb,HarlandLang:2012qz} 
the continuum background to the production of the $\chi_c(0^{+})$ state 
via two-body $\pi^{+} \pi^{-}$ and $K^{+} K^{-}$ decays was considered. 
For exclusive production of other mesons
see e.g. \cite{Lebiedowicz:2010yb, Cisek:2011vt, Lebiedowicz:2013vya},
where mainly the non-central processes were discussed.

\begin{figure}
(a)\includegraphics[width=3.8cm]{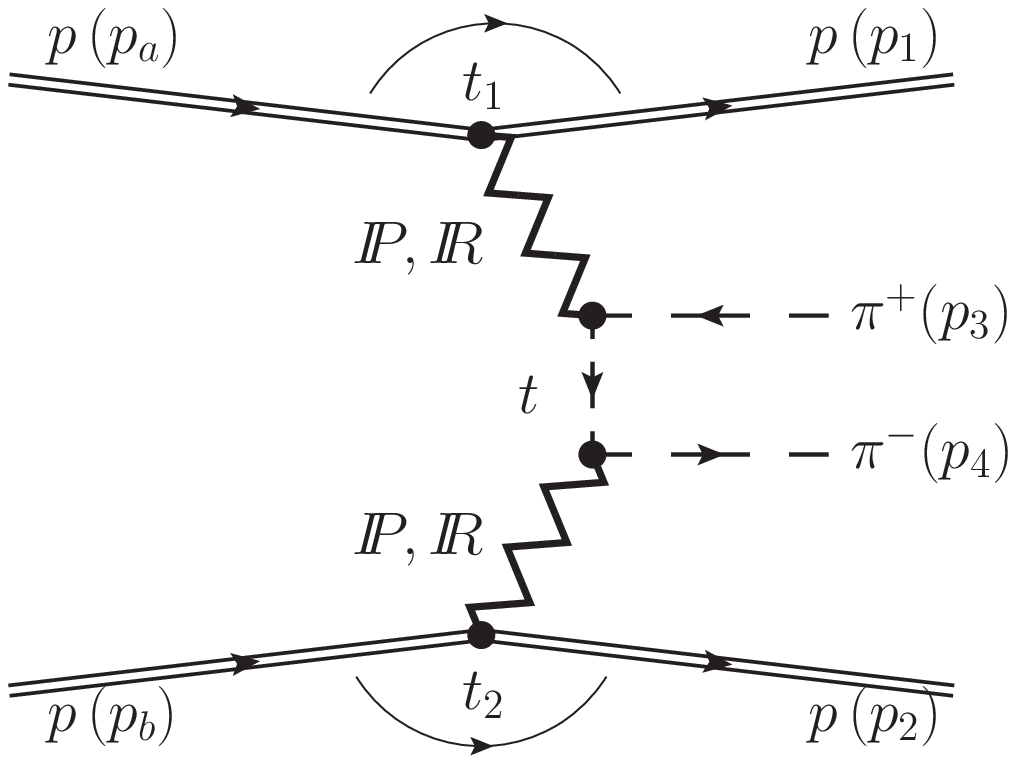}
   \includegraphics[width=3.8cm]{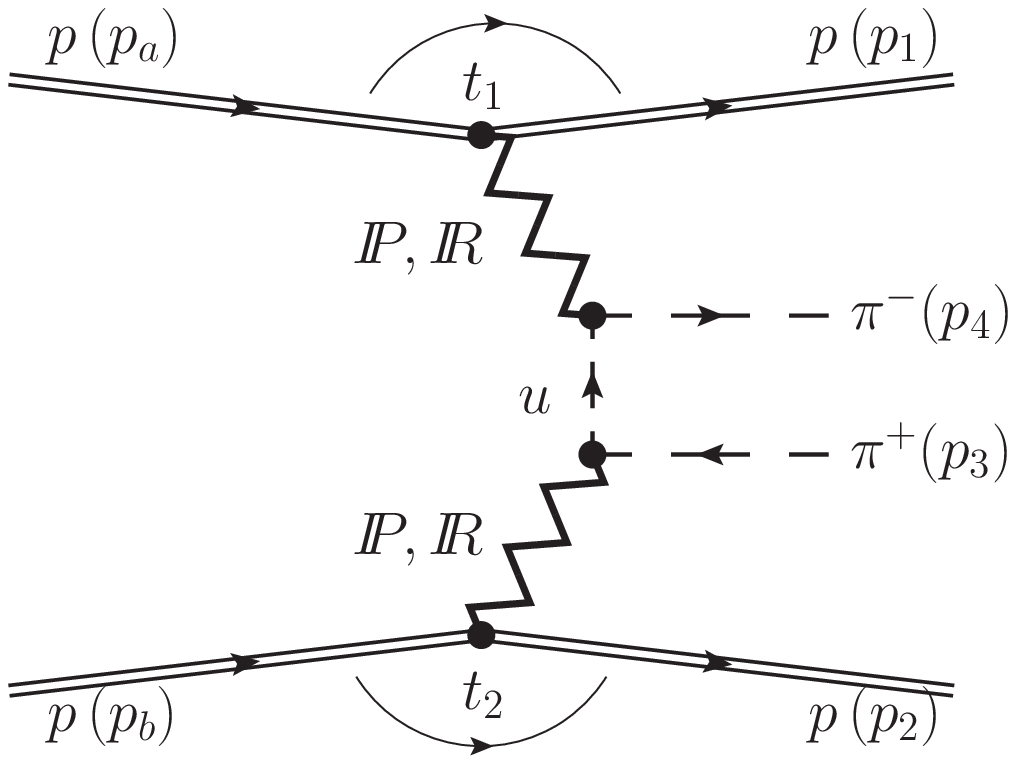}    
(b)\includegraphics[width=3.8cm]{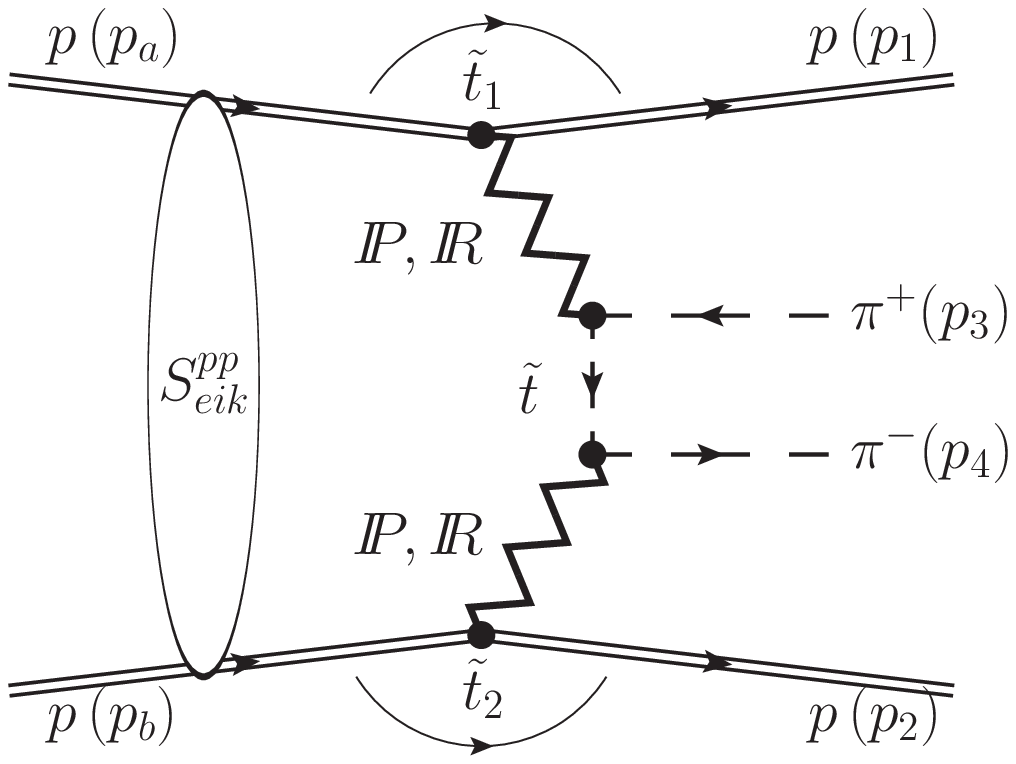}
   \includegraphics[width=3.8cm]{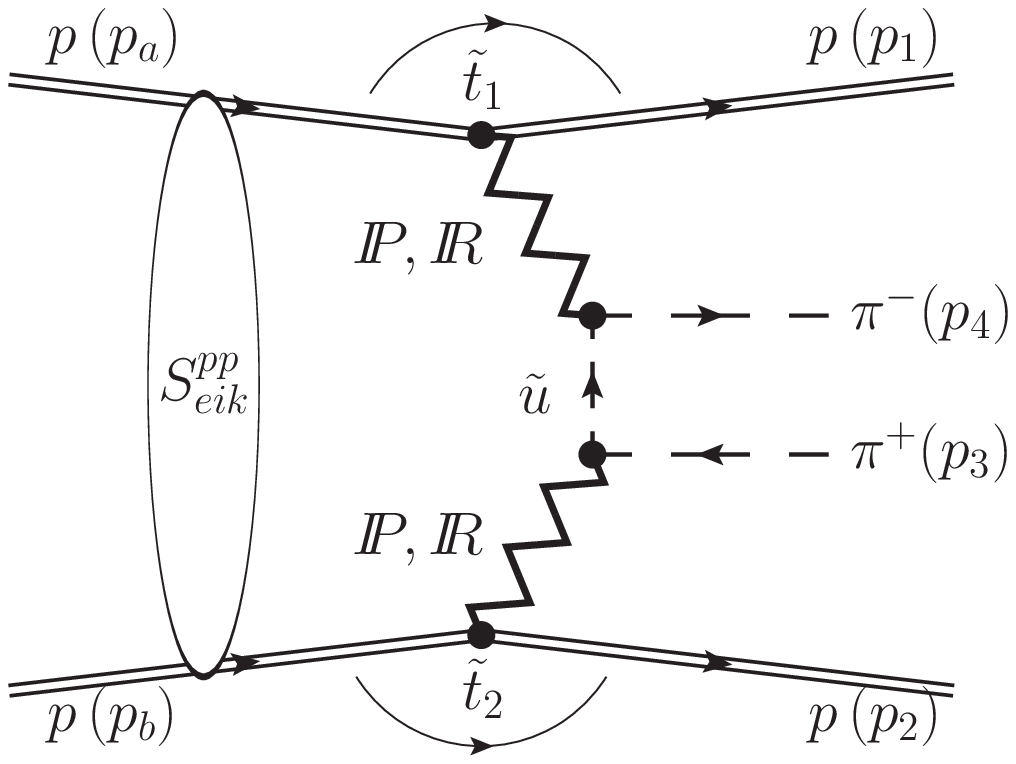}     
  \caption{\label{fig:diagrams_Born}
  \small (a) Born amplitudes for the $p p \to p p \pi^+ \pi^-$ process.
         (b) Absorptive correction amplitudes due to the $pp$ interaction.}
\end{figure}

Some time ago we proposed a simple phenomenological model 
for the $\pi^{+} \pi^{-}$-continuum mechanism
[see diagrams in Fig.~\ref{fig:diagrams_Born}~(a)]
using the tools of Regge theory \cite{Lebiedowicz:2009pj},
where perturbative QCD cannot be reliably applied.
In this model the parameters of pomeron and subleading reggeon exchanges
were adjusted to describe total and elastic $\pi N$ scattering.
We expect that our parametrization correctly extrapolate the interaction
parameters to higher energies where there are no experimental data.
The non-resonant model can be supplemented to include 
the $p p$ or $p \bar{p}$ absorption effects \cite{Lebiedowicz:2011nb,Lebiedowicz:thesis}; 
see diagrams in Fig.~\ref{fig:diagrams_Born}~(b).
The largest uncertainties in the Lebiedowicz-Szczurek model are due
to the unknown off-shell pion form-factor and the absorption corrections 
(the soft survival factor due to screening corrections).
The absorption is done in the eikonal approximation. 
The absorption effects lead to substantial damping of the cross section. 
The damping depends on the collision energy and kinematical variables.
The discussed here model, with reasonable vertex form factors accounting for
off-shellness of non-piont-like pions in the middle of diagrams 
in Fig.~\ref{fig:diagrams_Born},
gives a rough description of the ISR data \cite{Waldi:1983sc,Breakstone:1989ty,Breakstone:1990at}.
To get a reasonable description of existing so far experimental data 
parameter(s) of the form factors has (have) to be adjusted 
\cite{Lebiedowicz:2011nb,Lebiedowicz:thesis}.

It is not clear if the considered so far absorption effects are sufficient
to describe the data. 
Any interaction between participating particles 
potentially leads to absorptive effects, as it destroys exclusivity of the process.
In Fig.~\ref{fig:diagrams_Born}~(b) we show schematically
absorptive amplitudes due to $\pi N$ interaction in the final state. 
Not all combination of interactions are shown in the figure. 
Some absorptive effects are included inherently in our calculation;
by using effective interaction fitted to describe $\pi N$ elastic scattering data.
\begin{figure}
  \includegraphics[width=4.cm]{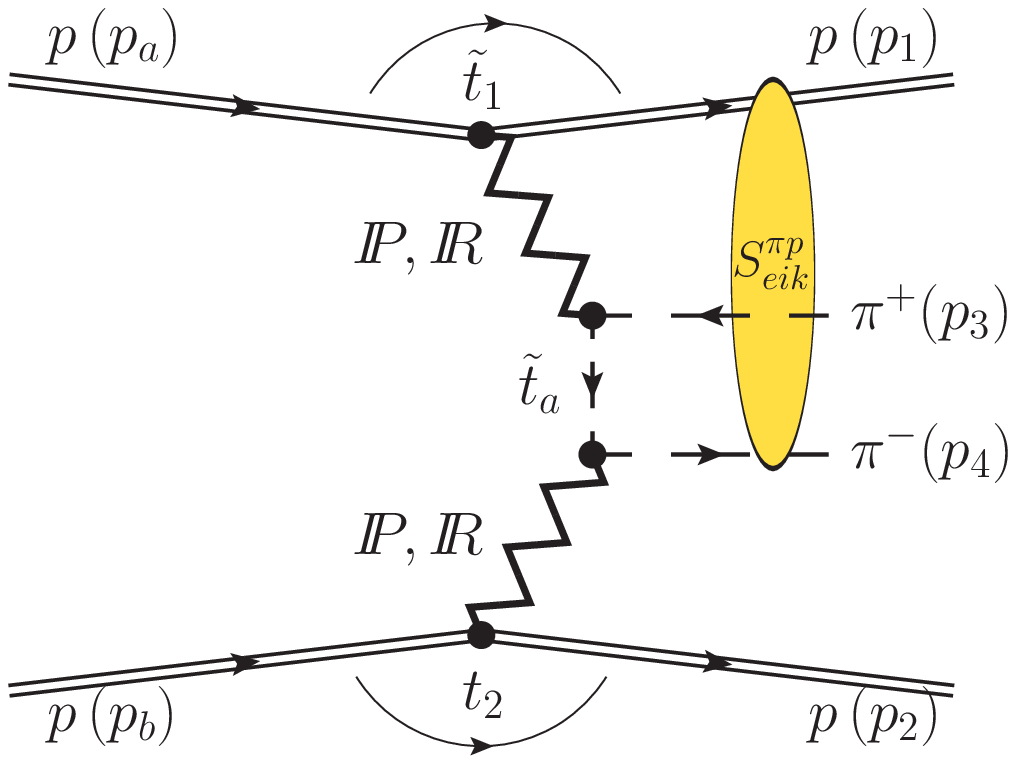}   
  \includegraphics[width=4.cm]{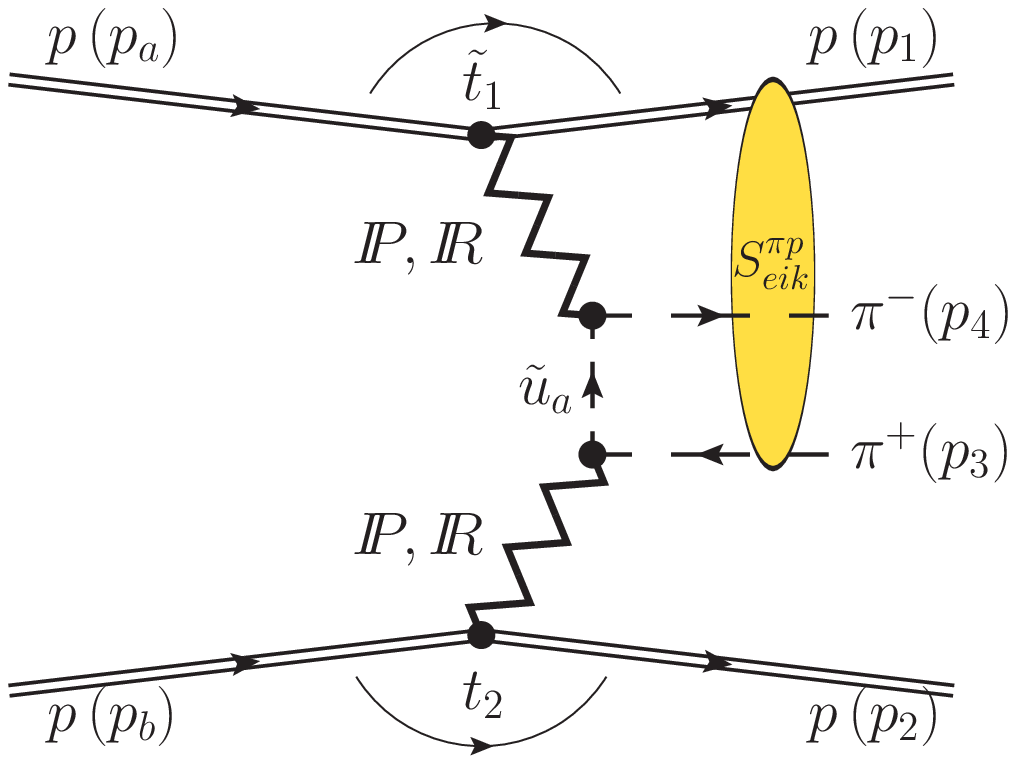}  
  \includegraphics[width=4.cm]{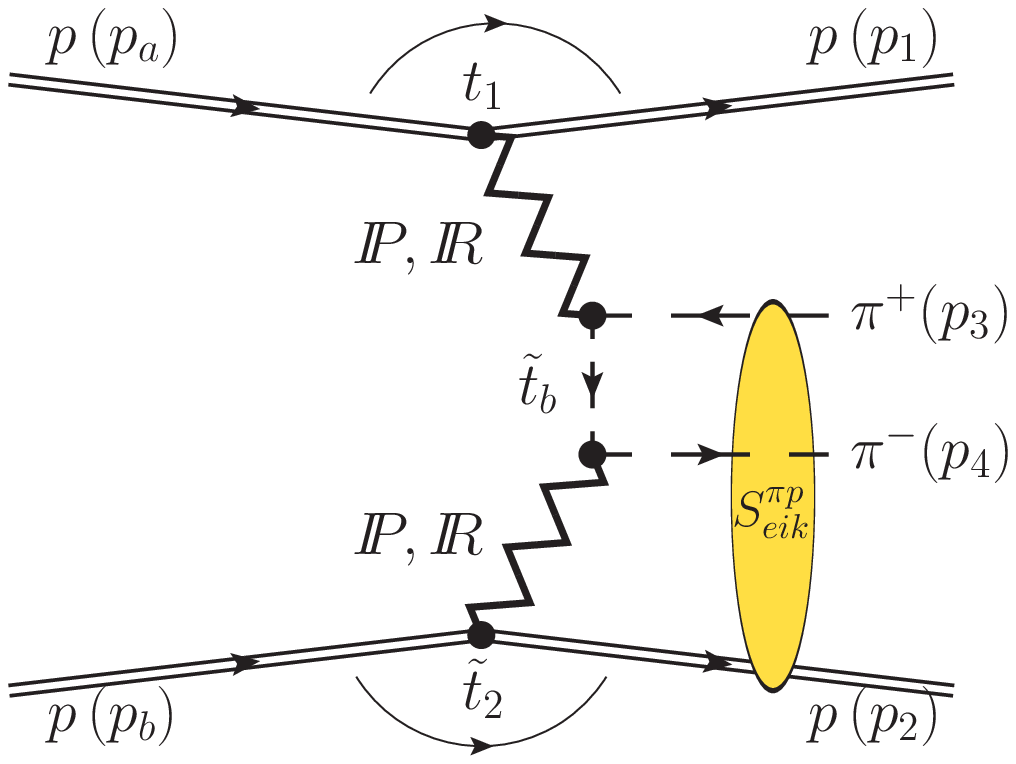}     
  \includegraphics[width=4.cm]{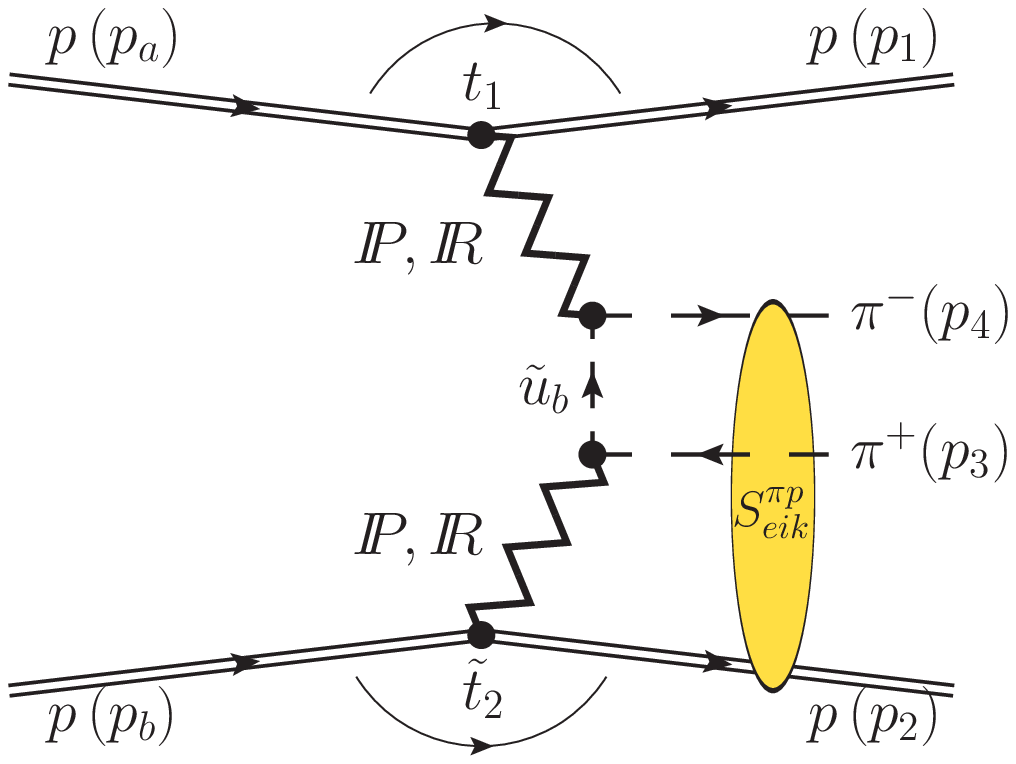}          
  \caption{\label{fig:diagrams_pip_absorption}
  \small New absorptive correction amplitudes for 
the $p p \to p p \pi^+ \pi^-$ process due to the $\pi p$ interaction
in the final state included in the present analysis.}
\end{figure}

Recently, the Lebiedowicz-Szczurek model \cite{Lebiedowicz:2009pj}
was implemented in GenEx MC \cite{Kycia:2014hea}.
There absorptive corrections are not taken into account explicitly.
The authors of \cite{Harland-Lang:2013dia} constructed a DIME Monte Carlo code,
where almost the same approach was implemented.
In \cite{Harland-Lang:2013dia} a two-channel eikonal model was considered.

In the present paper we will include the additional absorptive effects,
not considered so far in the literature, and quantify their role for 
total cross section and for many differential distributions for the considered process.
Our theoretical results will be compared to recent experimental results
obtained by the STAR \cite{Adamczyk:2014ofa} 
and CDF \cite{Aaltonen:2015uva,Albrow_Project_new} collaborations.
We will also show some predictions for ALICE, ATLAS and CMS experiments.

\section{Born amplitude}
%
%

The amplitude squared for the $p p \to p p \pi^+ \pi^-$ process 
(with four-momenta $p_a + p_b \to p_1 + p_2 + p_3 + p_4$) 
considered within the framework of Regge theory 
with the central $\pi^{+}\pi^{-}$ system 
produced by the exchange of two pomeron/reggeon in the $t$-channel, 
as shown in Fig.~\ref{fig:diagrams_Born}, can be written as
\begin{eqnarray}
|\mathcal{M}|^{2} = 
|\mathcal{M}_{I=0}|^{2} +
|\mathcal{M}_{I=1}|^{2} + 
|\mathcal{M}_{I=2}|^{2} \,,
\label{Regge_amplitude_full}
\end{eqnarray}
where the isospin amplitudes can be decomposed to the Regge ingredients as
\begin{eqnarray}
&&\mathcal{M}_{I=0} =\mathcal{M}^{\Pom \Pom} + 
                     \mathcal{M}^{\Pom f_{2 \Reg}} + 
                     \mathcal{M}^{f_{2 \Reg} \Pom} +
                     \mathcal{M}^{f_{2 \Reg} f_{2 \Reg}} +
\Braket{1,0;1,0|0,0} \mathcal{M}^{\rho_{\Reg} \rho_{\Reg}} \,, 
\label{amp_0}\\
&&\mathcal{M}_{I=1} =\mathcal{M}^{\Pom \rho_{\Reg}} + 
                     \mathcal{M}^{\rho_{\Reg} \Pom} + 
                     \mathcal{M}^{f_{2 \Reg} \rho_{\Reg}} +
                     \mathcal{M}^{\rho_{\Reg} f_{2 \Reg}} \,, 
\label{amp_1}\\
&&\mathcal{M}_{I=2} = 
\Braket{1,0;1,0|2,0} \mathcal{M}^{\rho_{\Reg} \rho_{\Reg}} \,.
\label{amp_2}
\end{eqnarray}
The Clebsch-Gordan coefficients $\Braket{j_{1},m_{1};j_{2},m_{2}|j,m}$ are 
\begin{eqnarray}
\Braket{1,0;1,0|0,0} = \sqrt{2/3} 
\; \; \; \mathrm{and} \; \; \;
\Braket{1,0;1,0|2,0} = -\sqrt{1/3}\,. \nonumber
\end{eqnarray}
The situation can be summarized as
\begin{eqnarray}
|\mathcal{M}_{I=0}|^{2} \gg
|\mathcal{M}_{I=1}|^{2} \gg
|\mathcal{M}_{I=2}|^{2} \,.
\end{eqnarray}
For the dominant pomeron-pomeron contribution we have $C$-parity $C = +1$ and isospin $I = 0$
of the produced $\pi^{+} \pi^{-}$ system.
In general, not only leading double pomeron exchanges contribute,
but also the subleading $f_{2 \Reg}$ ($C = +1$) and $\rho_{\Reg}$ ($C = -1$) 
reggeon exchanges.
\footnote{
The $\rho_{\Reg} \rho_{\Reg}$ component is negligible
(see the strength parameters in Table~2.1 of \cite{Lebiedowicz:thesis}) 
and was omitted in our analysis.
We emphasize, that at lower energies (COMPASS, ISR) 
the subleading $f_{2 \Reg}$ exchanges constitute a large contribution 
to the total cross section and must be included in addition to the pomeron exchanges;
see e.g. section~2.3 of \cite{Lebiedowicz:thesis}.
Furthermore, there is a large interference effect 
between the different components in the amplitude of about 50$\%$
(the total cross section in full phase space),
see section~2.6.2 of \cite{Lebiedowicz:thesis}.
As we shall see in the results section 
imposing limitations on pion rapidity $|y_{\pi}| < 1$
and going to higher energies reduces the role of subleading $f_{2 \Reg}$ exchanges,
however, due to their non-negligible interference effects 
with the leading $\Pom \Pom$ term we keep them explicitly in our calculations.}

The Born amplitude with the intermediate pion exchange can be written as
\begin{eqnarray} 
\mathcal{M}_{pp \to pp \pi^+ \pi^-}^{\mathrm{Born}}&=&
M_{13}(s_{13},t_1)
\frac{F_{\pi}^{2}(t)}{t-m_{\pi}^{2}}
M_{24}(s_{24},t_2)
+
M_{14}(s_{14},t_1)
\frac{F_{\pi}^{2}(u)}{u-m_{\pi}^{2}}
M_{23}(s_{23},t_2)\,,
\label{Regge_amplitude}
\end{eqnarray}
where the subsystem amplitudes $M_{ij}(s_{ij},t_{i})$ 
denotes ``interaction'' between forward proton ($i=1$)
or backward proton ($i=2$) and one of the two pions
($j=3$ for $\pi^{+}$ or $j=4$ for $\pi^{-}$). 
The energy dependence of the $\pi p$ subsystem amplitudes $M_{ij}$ is parametrised
in terms of the pomeron and the $f_{2 \Reg}$ reggeon exchange
\begin{eqnarray}
M_{ij}(s_{ij},t_{i}) =&&
\eta_{\Pom} \, s_{ij} \, C_{\Pom}^{\pi N}\left( \frac{s_{ij}}{s_0} \right)^{\alpha_{\Pom}(t_{i})-1} 
\exp \left( \frac{B_{\Pom}^{\pi N}}{2} t_{i} \right) \nonumber \\
&&+ \eta_{f_{2 \Reg}} \, s_{ij} \, C_{f_{2 \Reg}}^{\pi N}\left( \frac{s_{ij}}{s_0} \right)^{\alpha_{f_{2 \Reg}}(t_{i})-1} 
\exp \left( \frac{B_{f_{2 \Reg}}^{\pi N}}{2} t_{i} \right)\,,
\label{amplitude_part}
\end{eqnarray}
where $\eta_{\Pom} = i$, $\eta_{f_{2 \Reg}} = -0.860895+i$,
$s_{ij}$ is the energy in the ($ij$) subsystem,
and the energy scale $s_{0}$ is fixed at $s_{0} = 1$~GeV$^2$.
The pomeron and reggeon trajectories, 
$\alpha_{\Pom}(t)$ and $\alpha_{f_{2 \Reg}}(t)$, respectively, 
are assumed to be of standard form, 
see for instance \cite{Donnachie:2002en}, that is, linear in $t$:
%
\begin{eqnarray}
&&\alpha_{\Pom}(t) = \alpha_{\Pom}(0)+\alpha'_{\Pom}\,t \,, \quad \qquad
  \alpha_{\Pom}(0) = 1.0808, \; \alpha'_{\Pom} = 0.25 \; \mathrm{GeV}^{-2}\,, \\
&&\alpha_{f_{2 \Reg}}(t) = \alpha_{f_{2 \Reg}}(0)+\alpha'_{f_{2 \Reg}}\,t \,, \quad \;
  \alpha_{f_{2 \Reg}}(0) = 0.5475, \; \alpha'_{f_{2 \Reg}} = 0.93 \; \mathrm{GeV}^{-2}\,.
\label{trajectories}
\end{eqnarray}
%
%
%
We found the slope parameters $B_{\Pom/f_{2 \Reg}}^{\pi N}$
from fitting the elastic $\pi^{+} p$ and $\pi^{-} p$ cross sections
\begin{eqnarray}
B_{\Pom}^{\pi N} = 5.5 \;\mathrm{GeV}^{-2}\,, \quad
B_{f_{2 \Reg}}^{\pi N} = 4 \;\mathrm{GeV}^{-2} \,.
\label{MN_slope_parameters}
\end{eqnarray}
Our model makes various simplifications,
but describes the data for elastic $\pi N$ scattering
fairly well for energies $\sqrt{s_{\pi N}} \gtrsim 2.5$~GeV
(see Fig.~2.2 of \cite{Lebiedowicz:thesis}).

\begin{figure}[!ht]
\includegraphics[width=0.48\textwidth]{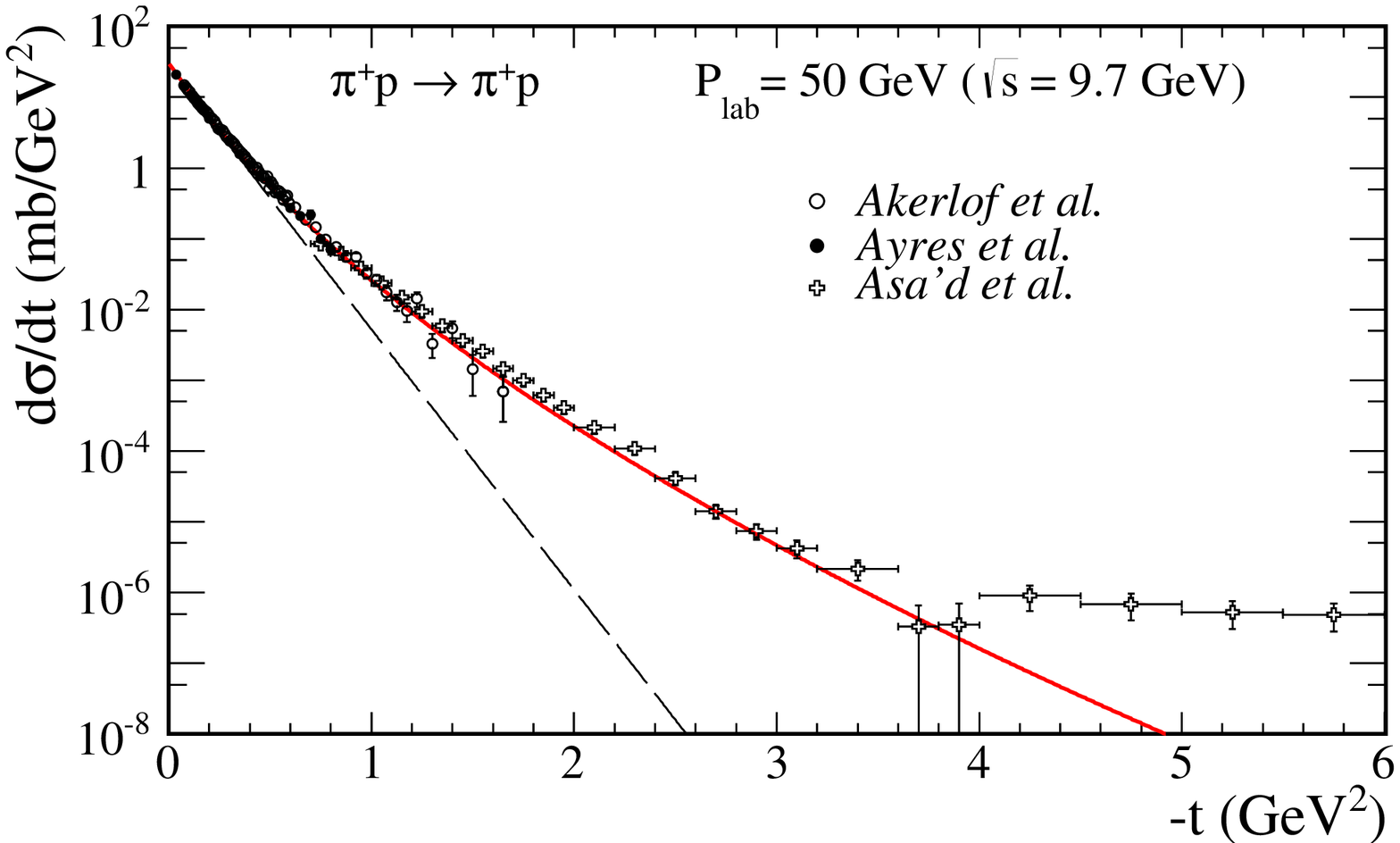}
\includegraphics[width=0.48\textwidth]{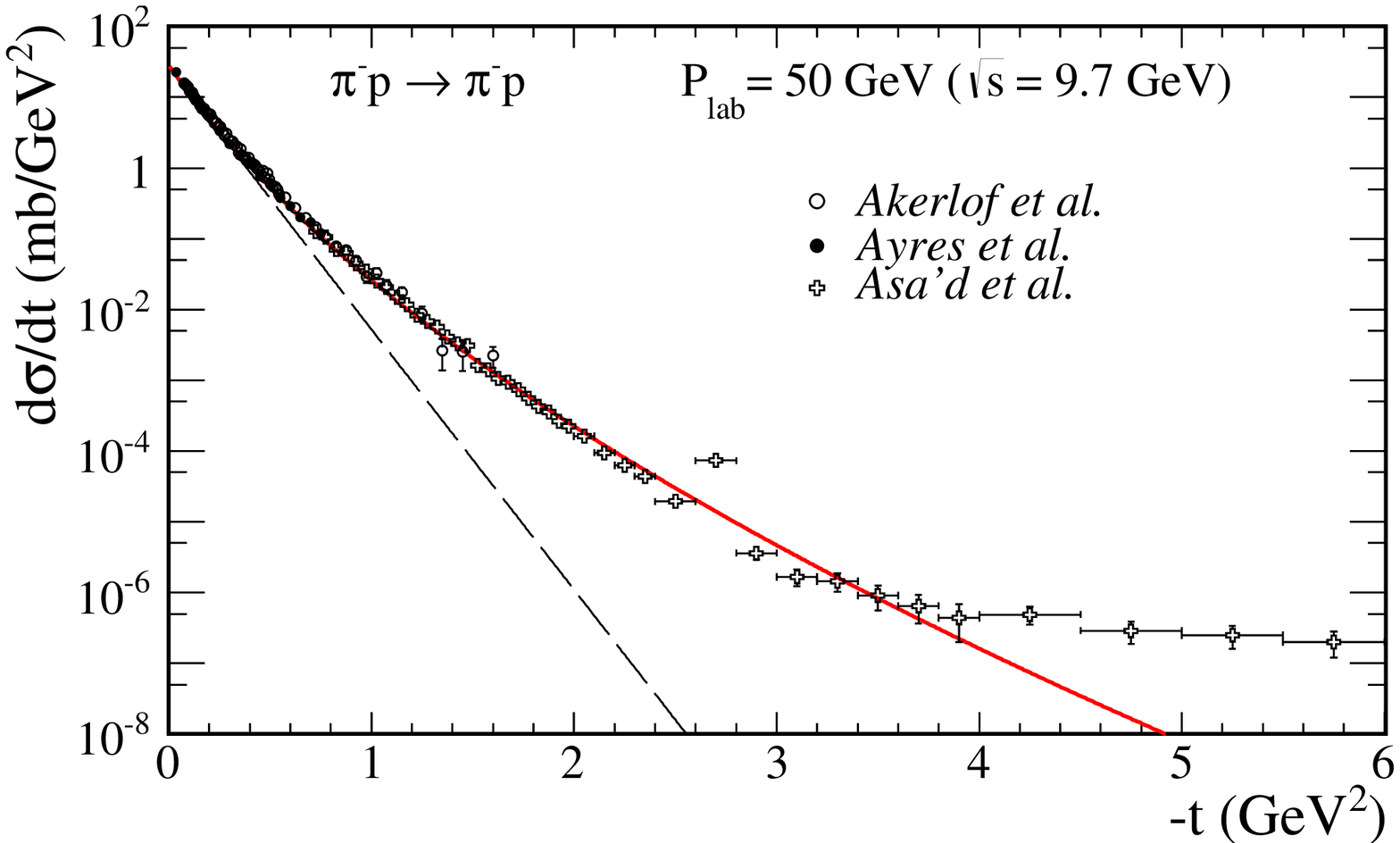}\\
\includegraphics[width=0.48\textwidth]{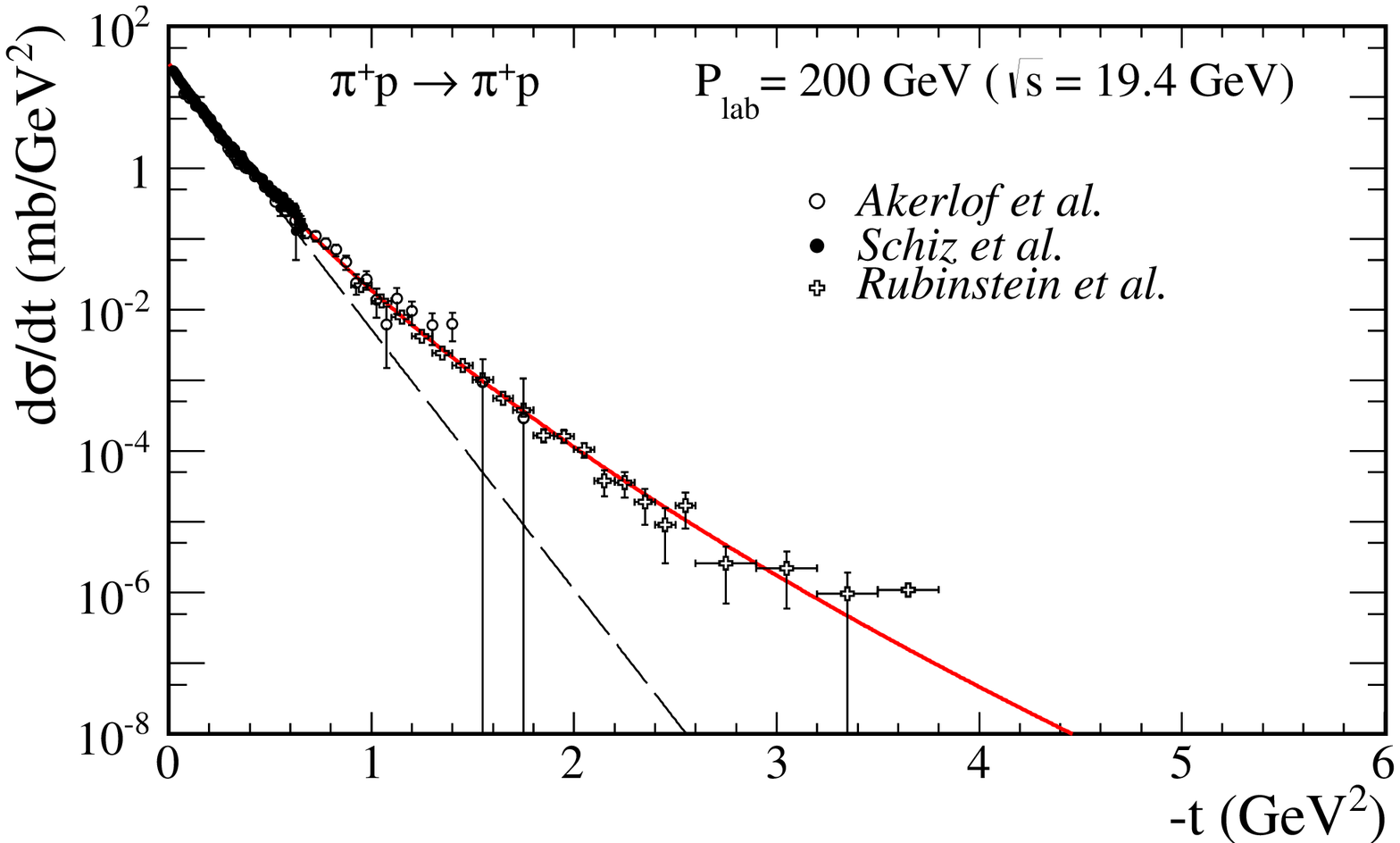}
\includegraphics[width=0.48\textwidth]{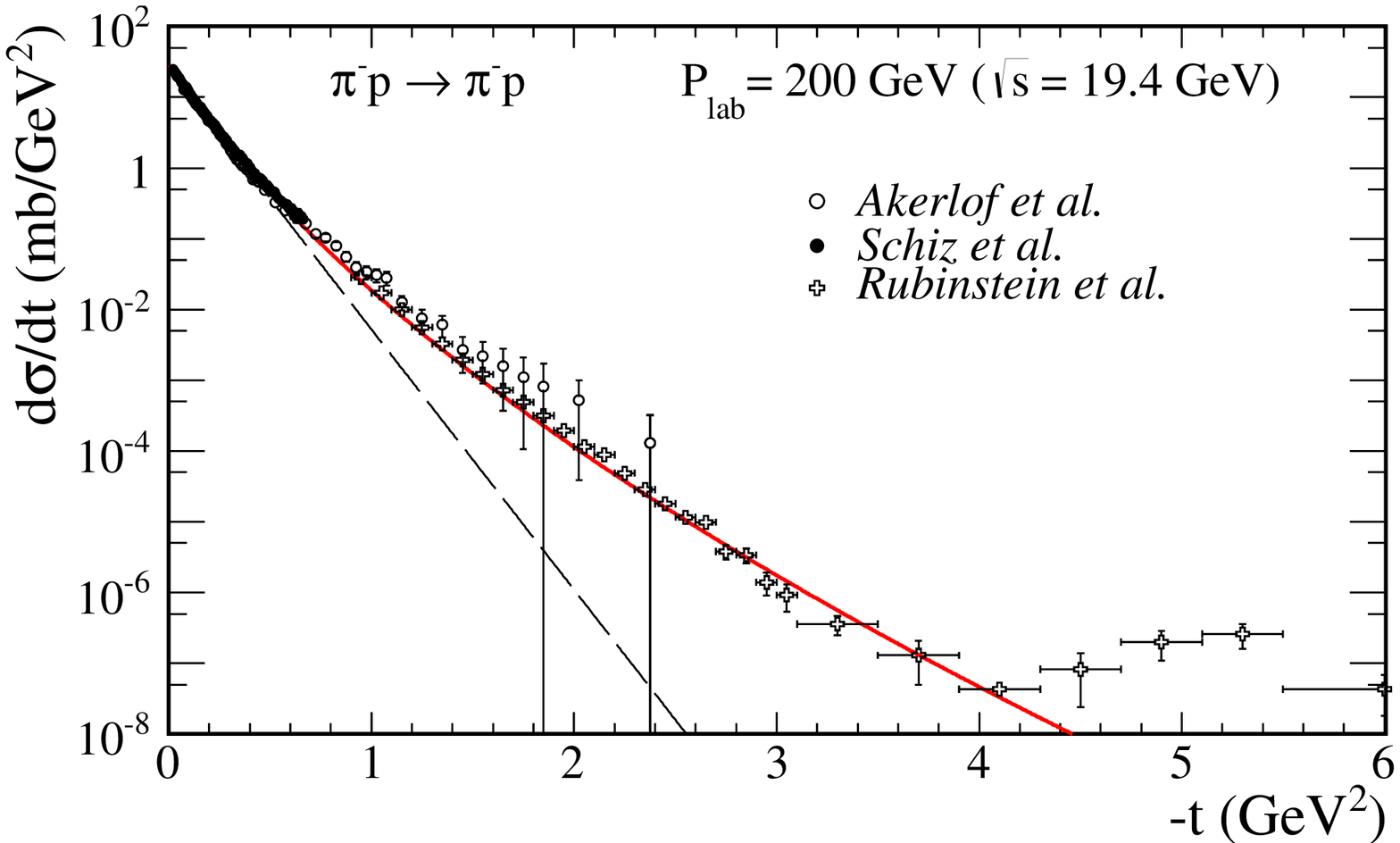}
  \caption{\label{fig:dsig_dt_elastic}
  \small
Differential distributions $d\sigma/dt$ for $\pi^{+}p$ (left panel)
and $\pi^{-}p$ (right panel) elastic scattering at incident beam momenta
$P_{lab} = 50$~GeV ($\sqrt{s} \simeq 9.7$~GeV) \cite{Akerlof:1976gk,Ayres:1976zm,Asad:1984my} and
$P_{lab} = 200$~GeV ($\sqrt{s} \simeq 19.4$~GeV) \cite{Akerlof:1976gk,Schiz:1979rh,Rubinstein:1984kf}.
The black dashed lines show results with formula (\ref{MN_slope_parameters})
while the red solid lines are obtained via the replacement (\ref{modified_t_dependence_orear}),
where $B_{0} = 6.5$~GeV$^{-2}$ and $\mu^{2} = 0.6$~GeV$^{2}$.
}
\end{figure}

So far we have assumed a simple exponential dependence of
the $\pi N$ subprocess amplitudes (\ref{amplitude_part}) 
which is valid only for small $|t|$ ($0.01 < -t < 0.4$~GeV$^{2}$).
At larger $|t|$ ($t_1$ or $t_2$ in the $2 \to 4$ case) the mechanism becomes more complicated. 
Here a subsequent exchange of two pomerons and 
the exchange of the pomeron together with the reggeon, 
or even pQCD effects (two-gluon exchange) may show up. 
\footnote{
Even description of the elastic $pp$ and $p \bar{p}$ scattering data is difficult.
In Ref.~\cite{Donnachie:2013xia} the authors
present a model including $\Pom$ + $\Pom \Pom$ + $ggg$ terms
and the linear pomeron trajectory.
Alternative approaches \cite{Ryskin:2011qe,Fazio:2013hza} combine the soft and hard pomeron exchanges
or the odderon exchange in addition \cite{Jenkovszky:2011hu}.
In the letter case the authors consider also a various forms of the non-linear pomeron trajectory.
In \cite{Khoze:2014nia} the role of the eikonalization 
of the $pp$ amplitude in both one- and the two-channel eikonal models were discussed.}
To give a more realistic $t$ dependence 
we suggest the following replacement (see also \cite{Cisek:2014ala})
%
%
%
\begin{eqnarray}
\exp\left( \frac{B_{\Pom}^{\pi N} }{2}\,t_i \right) 
\left( \frac{s_{ij}}{s_0} \right)^{\alpha'_{\Pom}t_{i}} \to &&
f(t_{i}, s_{ij}) = \exp \left( \mu^{2} B(s_{ij}) \right) 
\exp \left( - \mu^{2} B(s_{ij}) \sqrt{1-\frac{t_{i}}{\mu^{2}}} \right); \nonumber \\
&&\;
B(s_{ij}) = B_{0} + 2\alpha'_{\Pom} \ln\left(\frac{s_{ij}}{s_0}\right)\,,
\label{modified_t_dependence_orear}
\end{eqnarray}
%
%
where the free parameters $\mu$, $B_{0}$ has been adjusted to the $\pi N$ elastic scattering data, 
as illustrated in Fig.~\ref{fig:dsig_dt_elastic}.
The so-called 'stretched exponential' parametrization $f(t_{i}, s_{ij})$
coincides at low $|t|$ with the simple exponential form
while at larger $|t|$ features a harder tail.
This function is close to a parametrization $\sim \exp(-b p_{\perp})$ 
suggested by Orear \cite{Orear:1964zz} for elastic $pp$-scattering.
The $\pi p$ data show a diffraction dip at $-t \thicksim 4$~GeV$^{2}$.

The extra form factors $F_{\pi}(t)$ and $F_{\pi}(u)$, in Eq.~(\ref{Regge_amplitude}),
``correct'' for the off-shellnes of the intermediate pions.
The form of the form factor is unknown in particular at higher values of $t$ or $u$
and they are parametrised in two ways:
%
\begin{eqnarray} 
&&F_{\pi}(t)=
\exp\left(\frac{t-m_{\pi}^{2}}{\Lambda^{2}_{off,E}}\right) \,,
\label{off-shell_form_factors_exp} \\
&&F_{\pi}(t)=
\dfrac{\Lambda^{2}_{off,M} - m_{\pi}^{2}}{\Lambda^{2}_{off,M} - t} \,
\label{off-shell_form_factors_mon} 
\end{eqnarray}
and for $F_{\pi}(u)$ we have to replace $t \longleftrightarrow u$.
These form factors are normalized to unity on the pion-mass-shell $F_{\pi}(m_{\pi}^{2}) = 1$.
In general, the parameter $\Lambda_{off}$ is not known precisely but, 
in principle, could be fitted to the normalized experimental data.
How to extract the off-shell parameters will be discussed in the Result section.


\section{Absorption corrections}
\label{sec:absorption_corrections}

So far only absorption effects due to $pp$ ($p \bar p$) interactions
were included in the literature in the calculations of cross sections for 
the $p p \to p p \pi^+ \pi^-$ ($p \bar p \to p \bar p \pi^+ \pi^-$)
reactions \cite{Lebiedowicz:2011nb, Lebiedowicz:thesis}.
Here we wish to include also the absorption effects
due to strong nonperturbative interaction 
of charged pions and (anti)protons in the final state,
see corresponding diagrams in Fig.~\ref{fig:diagrams_pip_absorption}).
The absorption amplitude including the $\pi N$ interactions
can be written in a similar way as that in the case of 
$p p$ ($p \bar p$) interaction, i.e., in the eikonal form
\begin{eqnarray}
{\cal M}_{pp \to pp \pi^{+} \pi^{-}}^{\pi p-\mathrm{rescattering}} &\approx&
\frac{i}{16 \pi^2 s_{14}} \int d^2 k_t \;
{\cal M}_{pp \to pp \pi^+ \pi^-}^{\mathrm{Born}}(s,{\tilde t}_1,t_2,{\tilde t}_a)
{\cal M}_{\pi^- p \to \pi^- p}^{\Pom-\mathrm{exchange}}(s_{14},k_t^2)
\nonumber \\ 
&+&
\frac{i}{16 \pi^2 s_{13}} \int d^2 k_t \;
{\cal M}_{pp \to pp \pi^+ \pi^-}^{\mathrm{Born}}(s,{\tilde t}_1,t_2,{\tilde u}_a)
{\cal M}_{\pi^+ p \to \pi^+ p}^{\Pom-\mathrm{exchange}}(s_{13},k_t^2)
\nonumber \\ 
&+&
\frac{i}{16 \pi^2 s_{23}} \int d^2 k_t \;
{\cal M}_{pp \to pp \pi^+ \pi^-}^{\mathrm{Born}}(s,t_1,{\tilde t}_2,{\tilde t}_b)
{\cal M}_{\pi^+ p \to \pi^+ p}^{\Pom-\mathrm{exchange}}(s_{23},k_t^2)
\nonumber \\ 
&+&
\frac{i}{16 \pi^2 s_{24}} \int d^2 k_t \;
{\cal M}_{pp \to pp \pi^+ \pi^-}^{\mathrm{Born}}(s,t_1,{\tilde t}_2,{\tilde u}_b)
{\cal M}_{\pi^- p \to \pi^- p}^{\Pom-\mathrm{exchange}}(s_{24},k_t^2)
\,. \qquad
\label{extra_corrections}
\end{eqnarray}
In formula (\ref{extra_corrections}) we have indicated
explicitly only crucial variables, 
mostly those arguments of ${\cal M}_{p p \to p p \pi^+ \pi^-}$
which get modified in comparison to the Born amplitude (\ref{Regge_amplitude}).
For example, the four-momenta squared of the Regge exchange
in the first stage of the interaction 
(see Fig.~\ref{fig:diagrams_pip_absorption}) get modified as
\begin{eqnarray}
{\tilde t}_1 = ({\tilde p}_1 - p_a)^2 \,, \quad
{\tilde t}_2 = ({\tilde p}_2 - p_b)^2 \,,
\end{eqnarray}
where the four-momenta of the intermediate nucleons are
${\tilde p}_1 = p_1 - k_t$ and ${\tilde p}_2 = p_2 - k_t$.
%
%
Here, we have introduced auxiliary four-vector
%
$k_t = (0,\vec{k}_t,0)$
%
to write formulas in a compact way.
Similarly, the modified four-momenta of pions being propagated
in the middle of the four-body $p p \to p p \pi^+ \pi^-$ subprocess 
can be calculated as
\begin{eqnarray}
&&{\tilde t}_a = (\tilde{q}_1-p_3)^2 \,, \quad {\tilde u}_a = (\tilde{q}_1-p_4)^2 \,, \nonumber \\
&&{\tilde t}_b = (\tilde{q}_2-p_4)^2 \,, \quad {\tilde u}_b = (\tilde{q}_2-p_3)^2 \,,
\label{modified_pion_propagator_parameter}
\end{eqnarray}
where ${\tilde q}_1 = p_a - {\tilde p}_1$ and ${\tilde q}_2 = p_b - {\tilde p}_2$ 
are the four-momenta of the (incoming) Regge exchanges.
We leave all other not explicitly indicated variables which appear 
in the Born amplitude(s) unchanged. This is an approximation
but sufficient for the purpose of the present first exploratory analysis.

The full amplitude includes all rescattering corrections
\begin{eqnarray}
{\cal {M}}_{pp \to pp \pi^{+} \pi^{-}} =
{\cal {M}}_{pp \to pp \pi^{+} \pi^{-}}^{\mathrm{Born}} + 
c_{pp} {\cal {M}}_{pp \to pp \pi^{+} \pi^{-}}^{pp-\mathrm{rescattering}} + 
c_{\pi p} {\cal {M}}_{pp \to pp \pi^{+} \pi^{-}}^{\pi p-\mathrm{rescattering}} \,.
\label{amp_full}
\end{eqnarray}
In principle the contributions due to the intermediate proton(s)
diffractive excitation(s) ($p \to N^{*}$) could be effectively included 
by increasing the prefactors.
In the present paper we shall take, however, $c_{pp}$ = $c_{\pi p}$ = 1.
\footnote{
How the extra multiplication of the absorption amplitude 
${\cal {M}}_{pp \to pp \pi^{+} \pi^{-}}^{pp-\mathrm{rescattering}}$
by a factor $c_{pp}=1.2$ modify the features of differential distributions
was shown in \cite{Lebiedowicz:thesis}, see, e.g., Figs.~2.48, 2.49, 2.50, and Table~2.5.}

In the next section we shall show effect of inclusion
of the extra absorption terms on total cross section 
as well as on differential distributions. 
They will lead to further decrease of the cross section 
for the $p p \to p p \pi^+ \pi^-$ or $p \bar p \to p \bar p \pi^+ \pi^-$ reactions. 
We expect that the effects may be very important when comparing 
results of our calculation with the recent STAR and CDF experimental data 
as well as with the forthcoming data of the ALICE, CMS, and ATLAS collaborations.

\section{Predictions for different experiments}
\label{sec:predictions}

In this section we shall present some selected results
for the discussed exclusive processes calculated for kinematic domains
relevant for the STAR, CDF, ALICE, CMS, and ATLAS experiments.
In particular, we wish to concentrate on the effect of the new absorption
corrections due to the pion-(anti)proton interaction.
We refer also the readers to section~2.6.2 of \cite{Lebiedowicz:thesis}
where only the absorptive corrections due to $NN$ interaction were discussed.

\begin{figure}[!ht]
\includegraphics[width=0.48\textwidth]{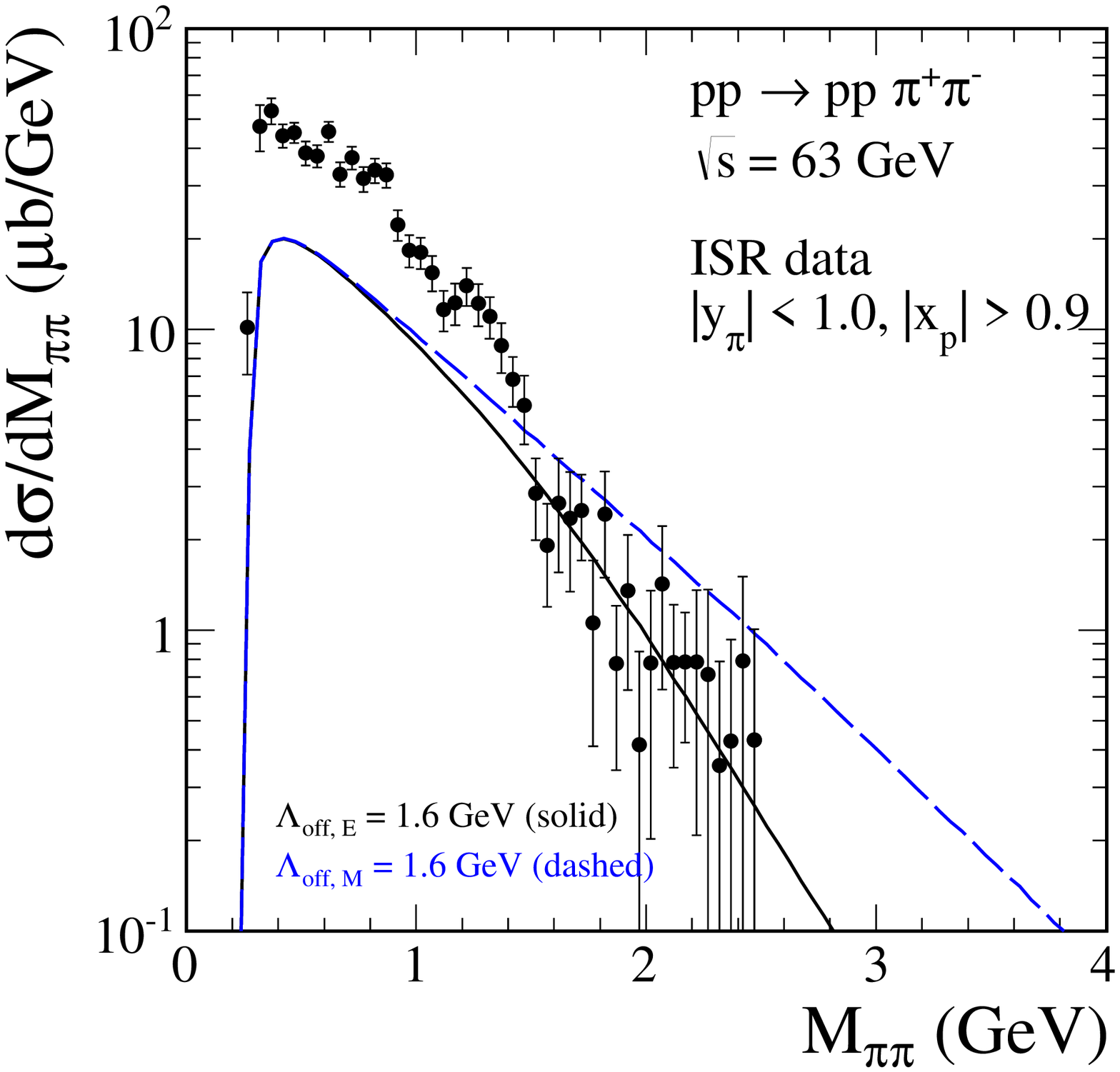}
\includegraphics[width=0.48\textwidth]{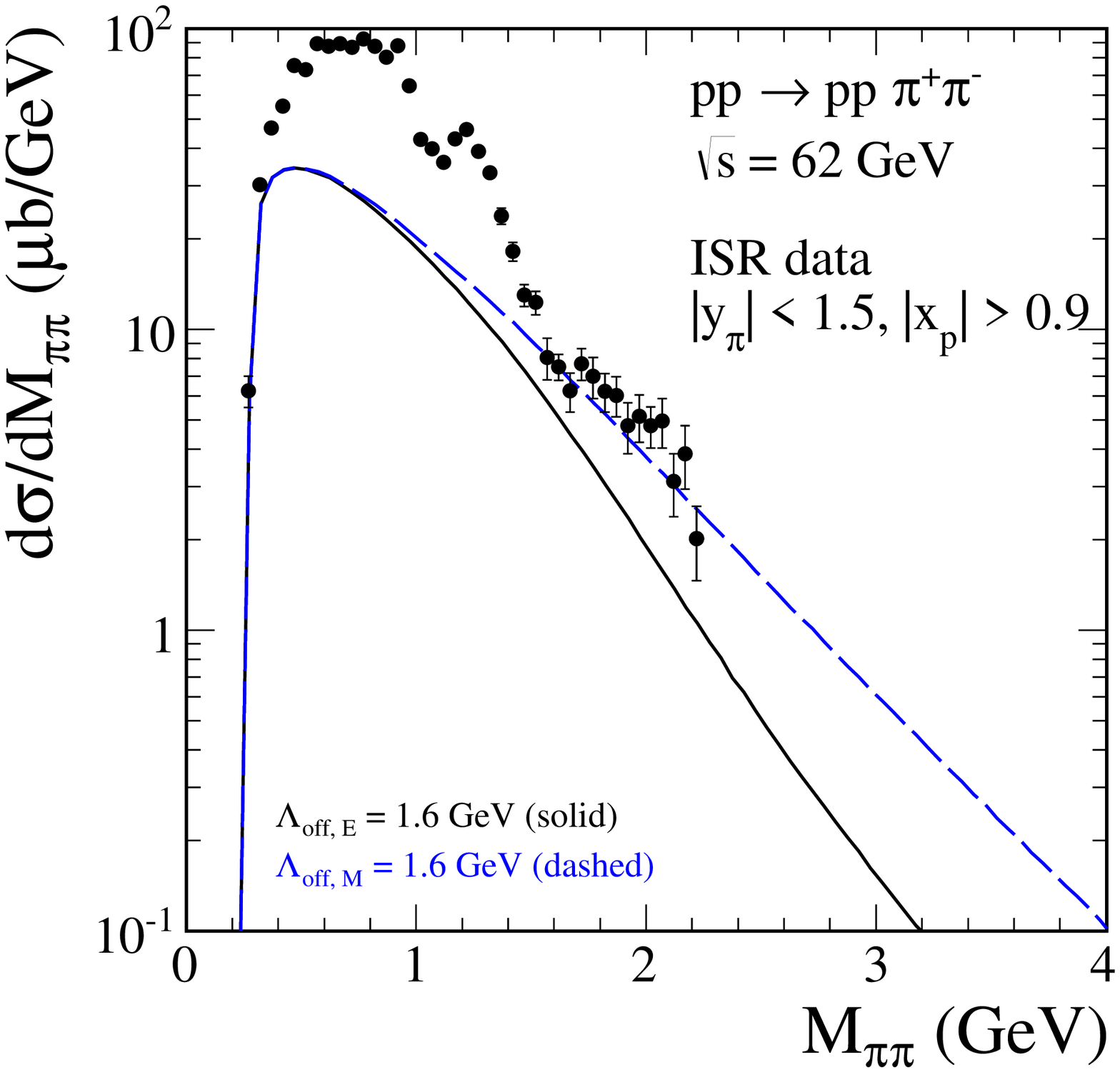}\\
\includegraphics[width=0.48\textwidth]{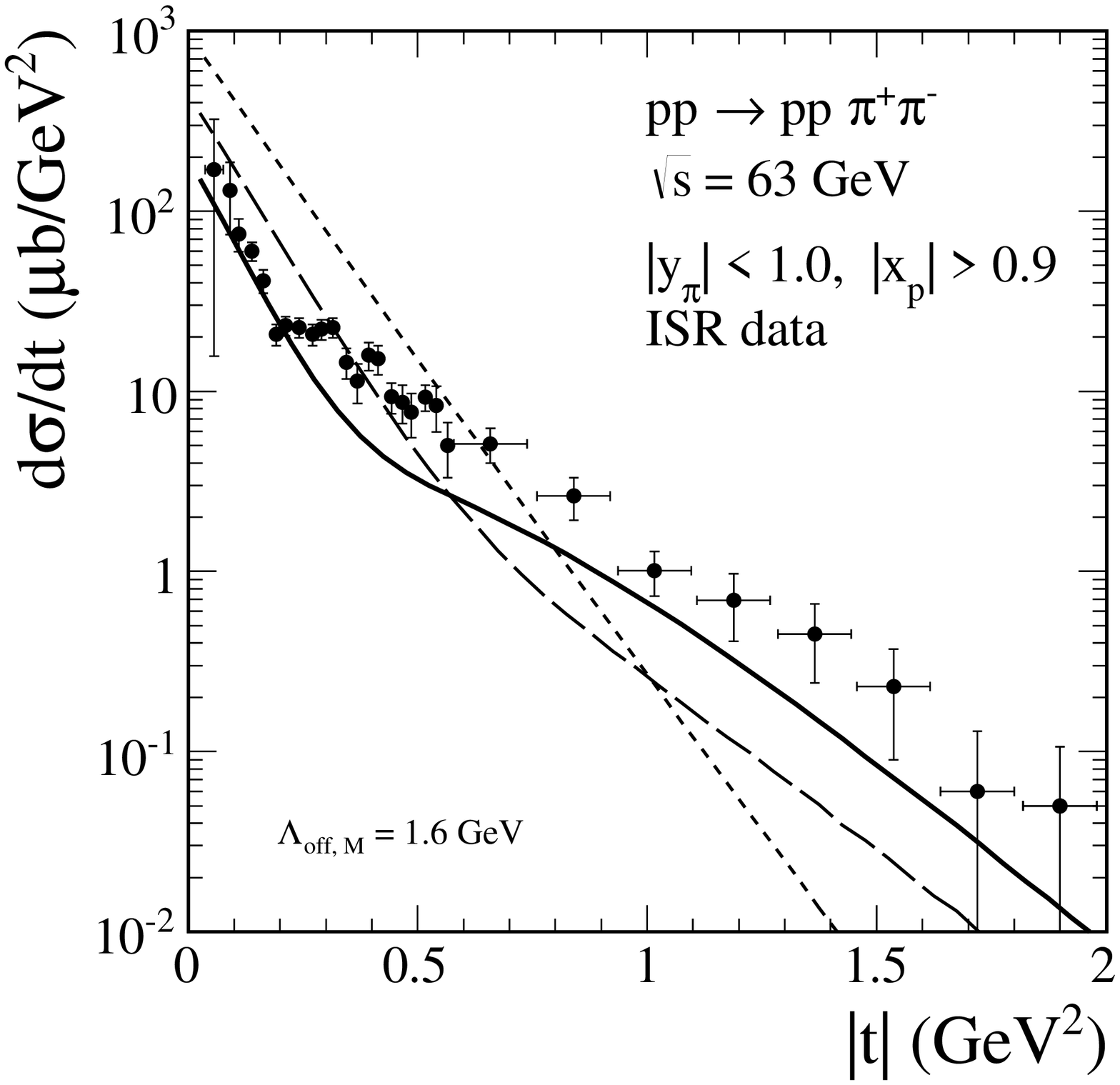}
\includegraphics[width=0.48\textwidth]{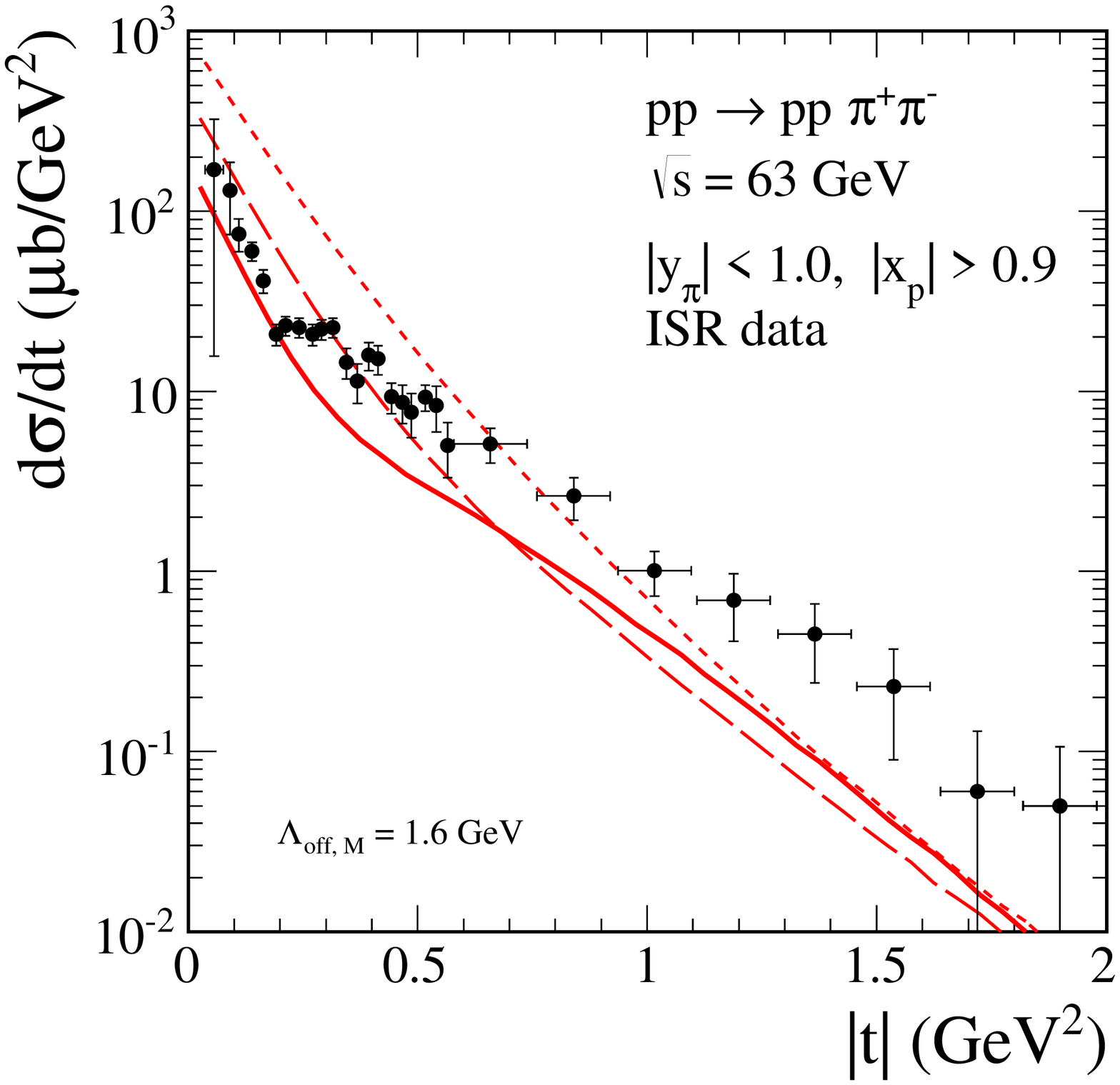}
  \caption{\label{fig:ISR}
  \small
Two-pion invariant mass distribution at ISR energies
with the ISR kinematical cuts indicated in the figure caption.
The ISR data \cite{Waldi:1983sc,Breakstone:1990at} are shown for comparison.
The blue dashed lines represent the results obtained 
for the monopole form factors 
[(\ref{off-shell_form_factors_mon}), $\Lambda_{off,M} = 1.6$~GeV], 
while the black solid lines are for the exponential form [(\ref{off-shell_form_factors_exp}), 
$\Lambda_{off,E} = 1.6$~GeV].
The bottom panels represent the $|t|$ distributions without 
(dotted lines), with the $pp$ absorption corrections (dashed lines),
and with all ($pp$ and $\pi N$) absorption corrections included (solid lines).
Results for the exponential (left bottom panel)
and for the 'stretched exponential' (right bottom panel) parametrizations of the $\pi p$ subsystem are shown.
}
\end{figure}

Before we go to the higher energies let us first 
discuss old ISR data \cite{Waldi:1983sc,Breakstone:1990at}.
In Fig.~\ref{fig:ISR} (top panels) we show results for two-pion invariant mass distributions.
The theoretical calculations including absorption corrections
have been compared with the ISR data.
In the calculation the form factor
for the off-shell pions was fixed as specified in the figure captions.
The choice of form factor leads to different behaviour at higher $M_{\pi \pi}$.
We also show (bottom panels) the result for the exponential 
and 'stretched exponential' $t$-dependences without and with absorption corrections.
The shape of the $t$ distributions is strongly modified by the absorption corrections
and is similar as obtained in the ISR experiment.

\subsection{STAR experiment}
\label{sec:STAR}

\begin{figure}[!ht]
\includegraphics[width=0.48\textwidth]{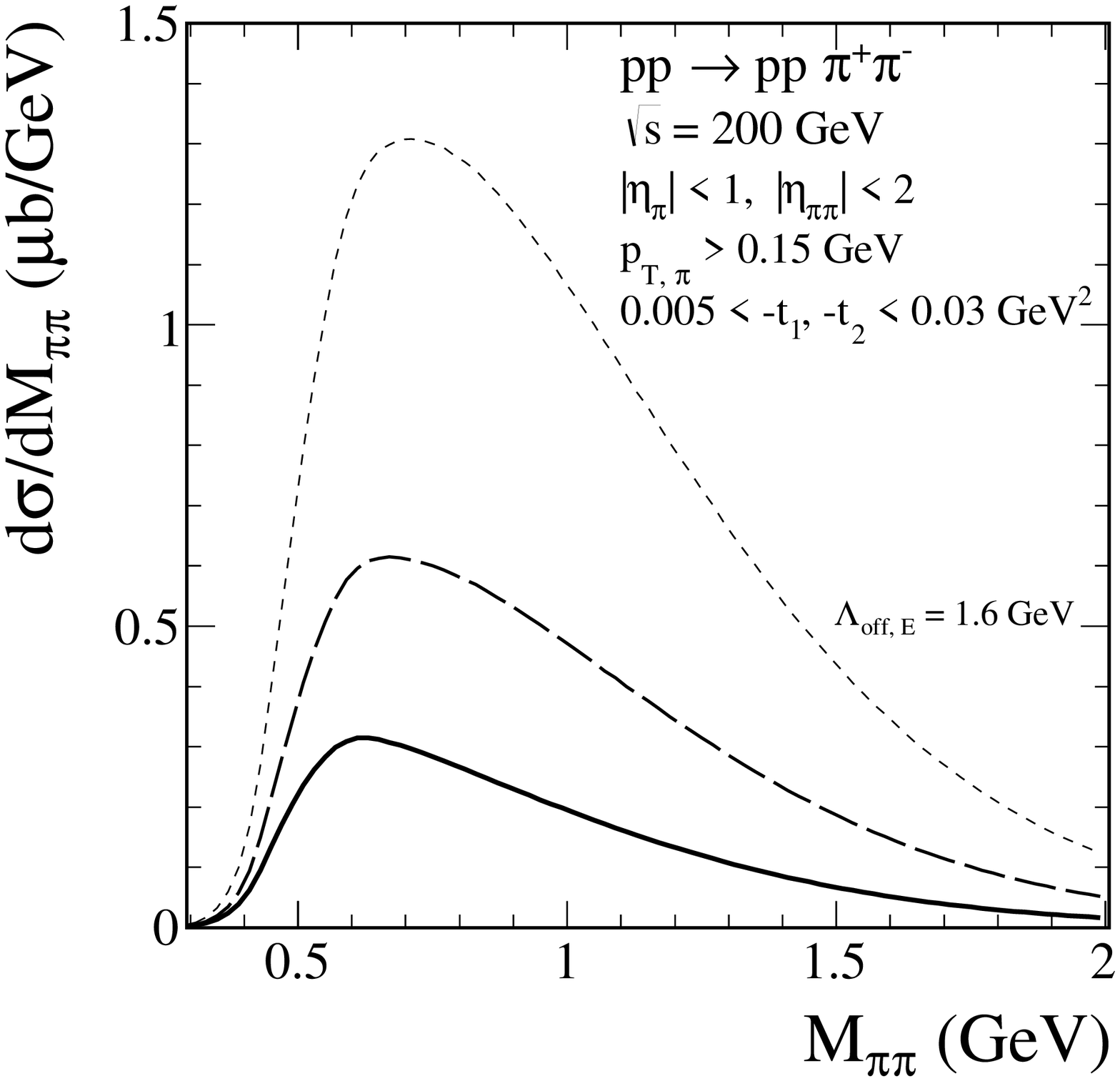}
\includegraphics[width=0.48\textwidth]{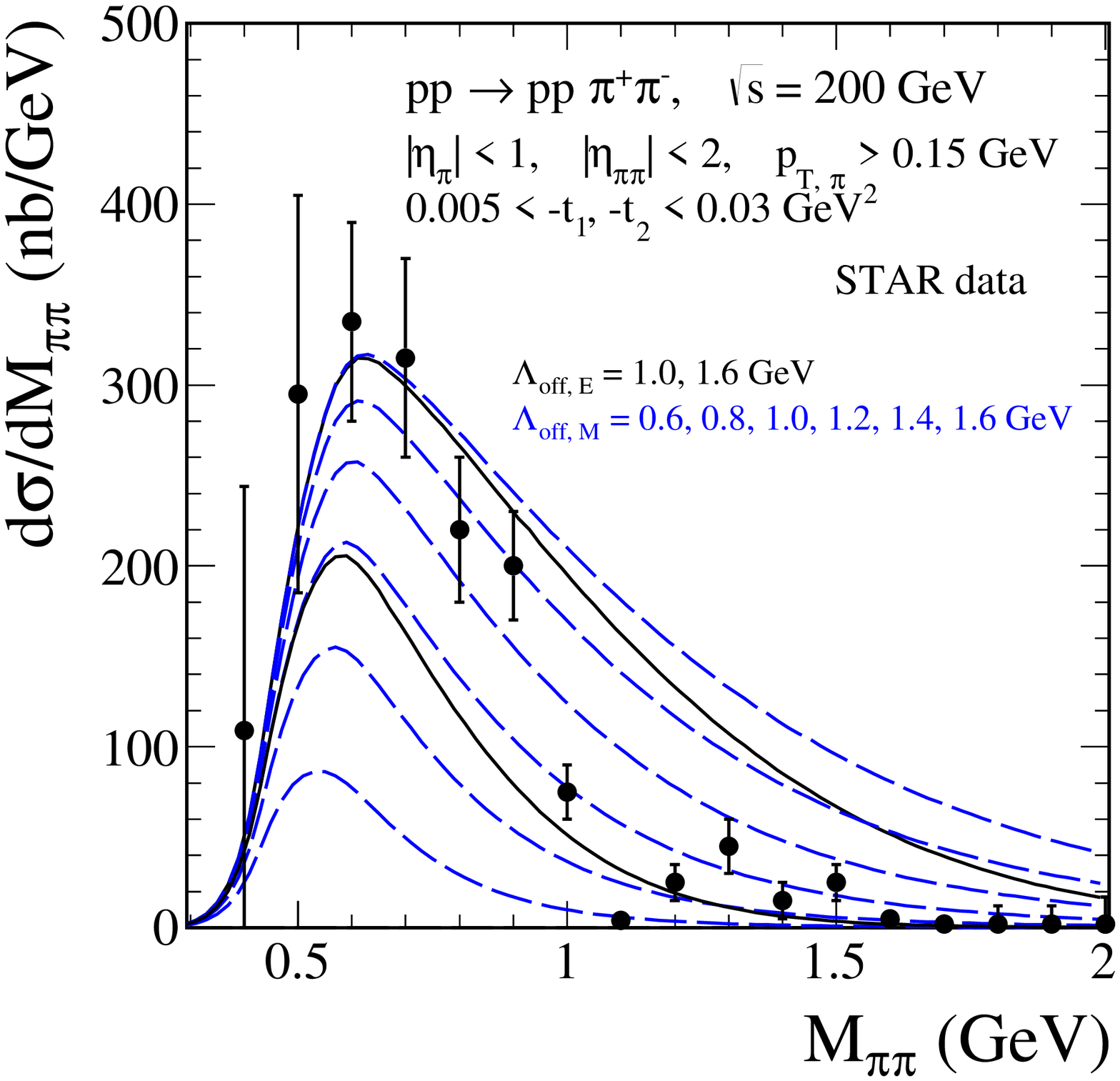}
  \caption{\label{fig:STAR_M34}
  \small
Two-pion invariant mass distribution at $\sqrt{s}=200$~GeV
with the STAR kinematical cuts specified in the figure caption.
The dotted line in the left panel corresponds to the Born calculation, 
the long-dashed and solid lines to calculations with the absorption 
effects due to the $pp$- and the $\pi p$-rescattering in addition, respectively.
In the right panel, the blue dashed lines represent the results 
with all absorption effects and obtained 
for the monopole form factors (\ref{off-shell_form_factors_mon})
for different choices of the cut-off parameter 
$\Lambda_{off,M} = 0.6$-$1.6$~GeV (from bottom to top).
The black solid lines are for the exponential form (\ref{off-shell_form_factors_exp})
and $\Lambda_{off,E} = 1.0$ and 1.6~GeV.
The STAR data \cite{Adamczyk:2014ofa} are shown for comparison.
}
\end{figure}

In Fig.~\ref{fig:STAR_M34} we present the invariant mass distributions
of the pion pair produced in the $p p \to p p \pi^+ \pi^-$ reaction
for the STAR kinematics ($\sqrt{s}=200$~GeV,
$|\eta_{\pi}| < 1$ and $p_{t, \pi} > 0.15$~GeV for both pions,
the pseudorapidity of the central $\pi^{+}\pi^{-}$ system $|\eta_{\pi\pi}| < 2$,
and in the four-momentum transfers range 0.005~$< -t_{1}, -t_{2} < 0.03$~GeV$^{2}$).
In the left panel we show result obtained in the Born approximation
(dotted line), the result when including proton-proton interactions 
(dashed line), and when including extra pion-nucleon interactions discussed 
in the present paper. 
We observe significant damping of the cross section as well as 
a small shift of the maximum towards smaller invariant masses. 
In the left panel we show result for different parameters of 
the off-shell form factors, together with the STAR experimental data.
One can observe that our predictions are quite sensitive 
to the form of the off-shell pion form factor
(\ref{off-shell_form_factors_exp}) or (\ref{off-shell_form_factors_mon})
and depend on the value of the cut-off parameters $\Lambda_{off}$.
If we describe the maximum of the cross section 
around $M_{\pi \pi} \sim$~0.6-0.7 GeV we overestimate the cross section
in the interval 1~$< M_{\pi \pi} <$~2~GeV essentially for both choices of
the form factor form. Part of the effect may be related to
an enhancement of the cross section due to $\pi \pi$ low-energy final
state interaction. This goes beyond the scope of the present paper
which concentrates on the new absorption effects.

\begin{figure}[!ht]
\includegraphics[width=0.48\textwidth]{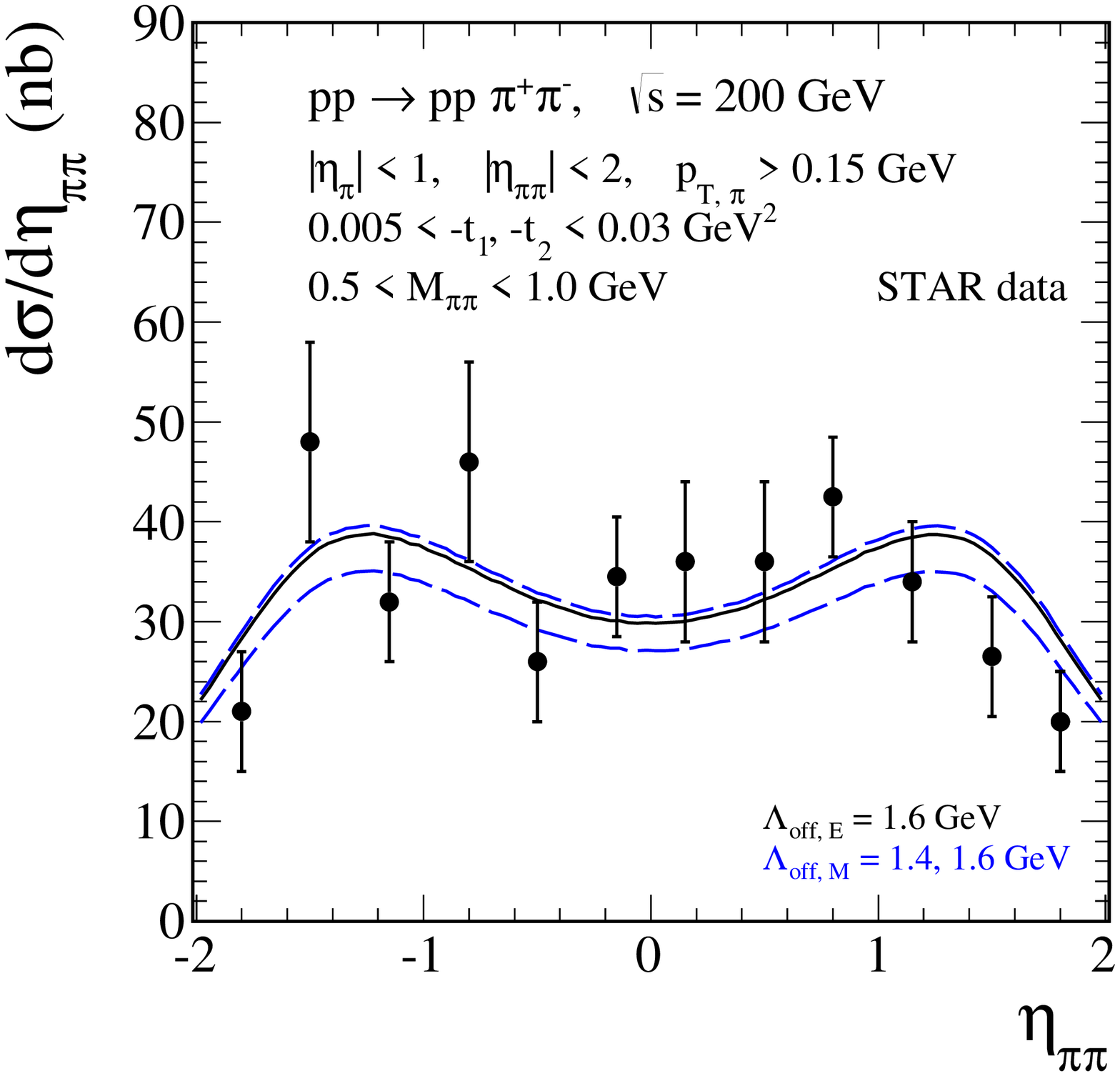}
\includegraphics[width=0.48\textwidth]{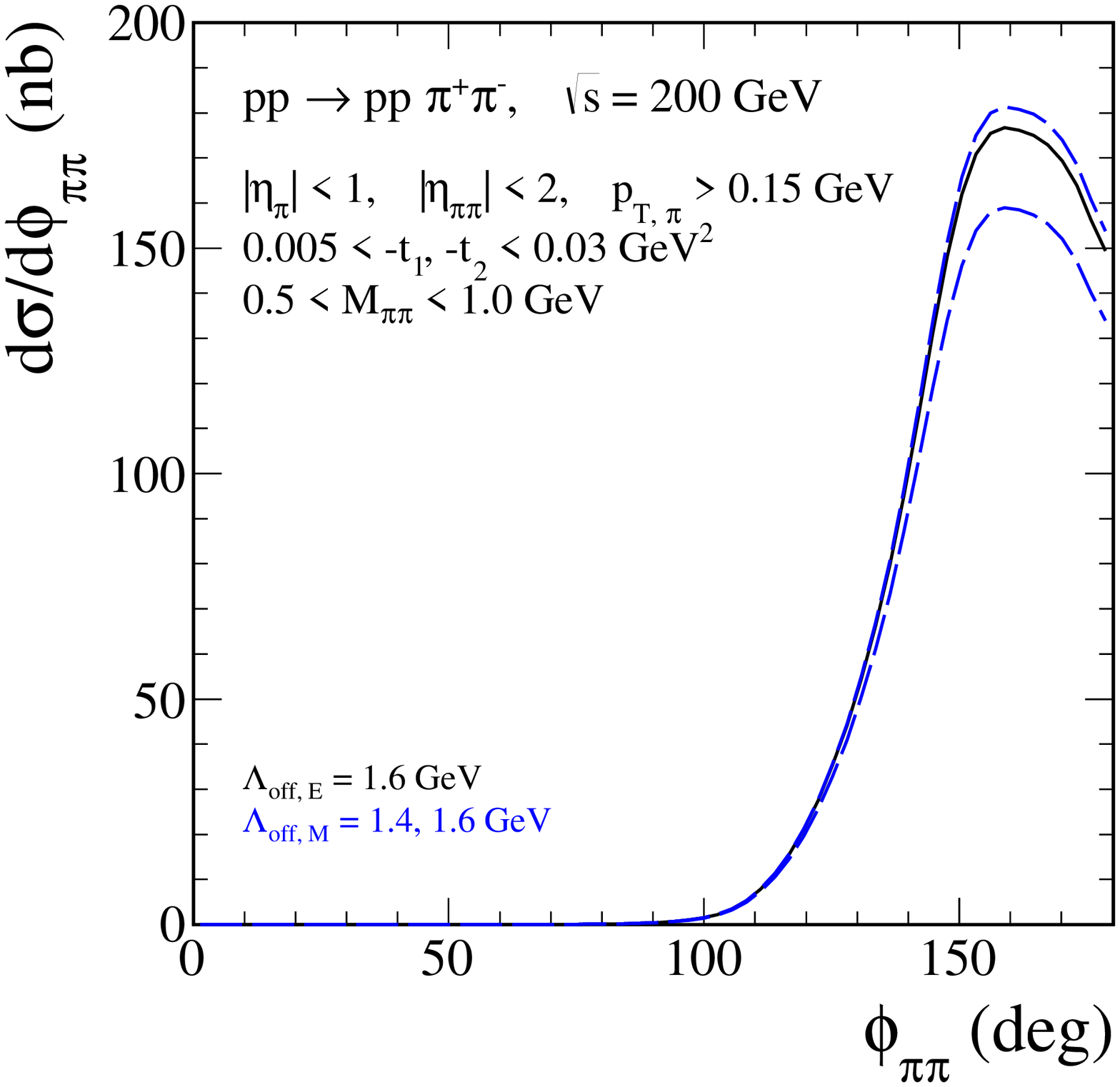}\\
\includegraphics[width=0.48\textwidth]{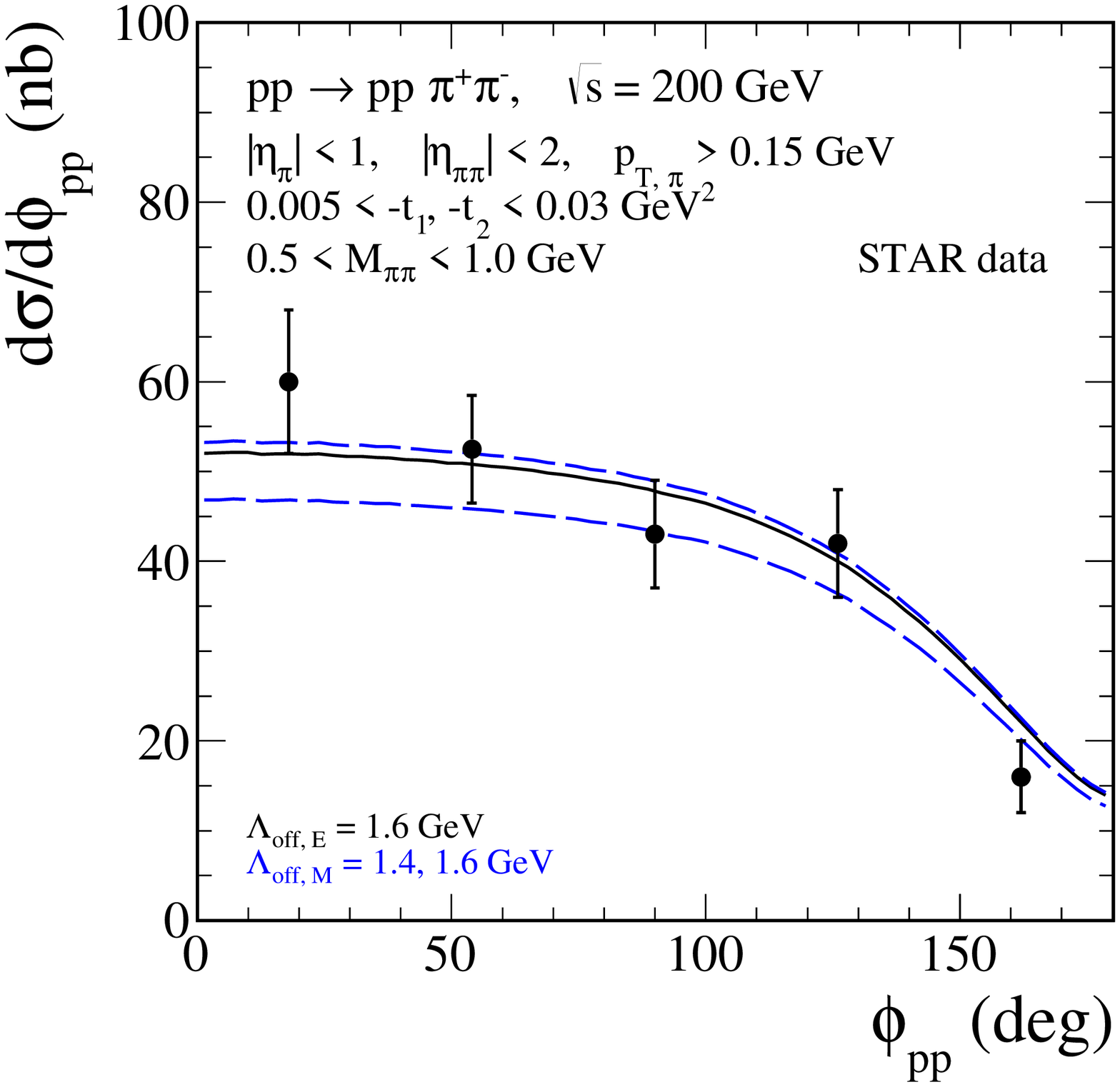}
\includegraphics[width=0.48\textwidth]{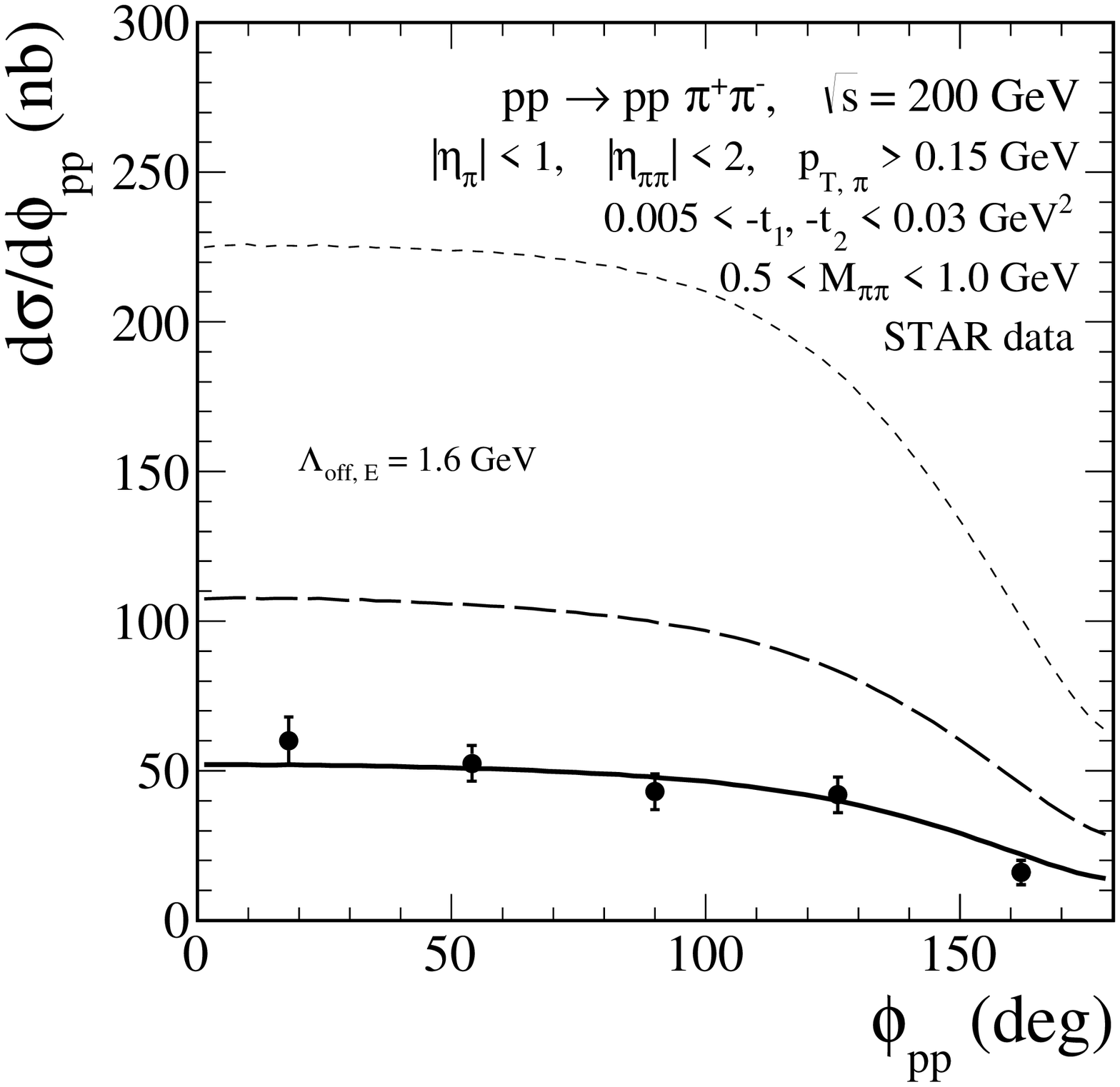}
  \caption{\label{fig:STAR_eta34}
  \small
The distributions in the pseudorapidity of the produced $\pi^{+}\pi^{-}$ system ($\eta_{\pi\pi}$)
and in the azimuthal angle between the outgoing pions ($\phi_{\pi \pi}$)
and between the outgoing protons ($\phi_{pp}$) at $\sqrt{s}=200$~GeV 
in the range of 0.5~$<M_{\pi\pi}<$~1.0~GeV.
In the right bottom panel we show the results without the absorption effects (the dotted line),
with the $pp$-rescattering (the long-dashed line),
and with the additional $\pi p$-rescattering (the solid line).
The STAR data \cite{Adamczyk:2014ofa} are shown for comparison.
}
\end{figure}

In Fig.~\ref{fig:STAR_eta34} (left panel) we show differential cross section
for the exclusive production of $\pi^{+}\pi^{-}$ system as 
a function of its pseudorapidity.
We conclude that in the range 0.5~$<M_{\pi\pi}<$~1.0~GeV 
both forms of the off-shell pion form factor
[(\ref{off-shell_form_factors_exp}) and (\ref{off-shell_form_factors_mon})] 
describe the data well for $\Lambda_{off} = 1.4 - 1.6$~GeV.
However, the agreement seems a bit misleading 
in the light of disagreement in the invariant mass distribution discussed above.
As will be discussed in this paper, the absorption effects 
usually strongly modify the distribution in relative azimuthal 
angle between the outgoing protons $\phi_{pp}$ and leave 
the shape of the $\phi_{\pi\pi}$ distribution almost unchanged. 
For the STAR (Phase I) visible kinematical range, 
that is at very small four-momentum transfers $|t|$,
one can observe only a damping of the cross section (the right bottom panel).
The decrease of $d\sigma/d\phi_{pp}$ and $d\sigma/d\phi_{\pi\pi}$
at $\phi \sim \pi$ is due to the condition $|\eta_{\pi\pi}| < 2$.

\subsection{CDF experiment}

\begin{figure}[!ht]
\includegraphics[width=0.48\textwidth]{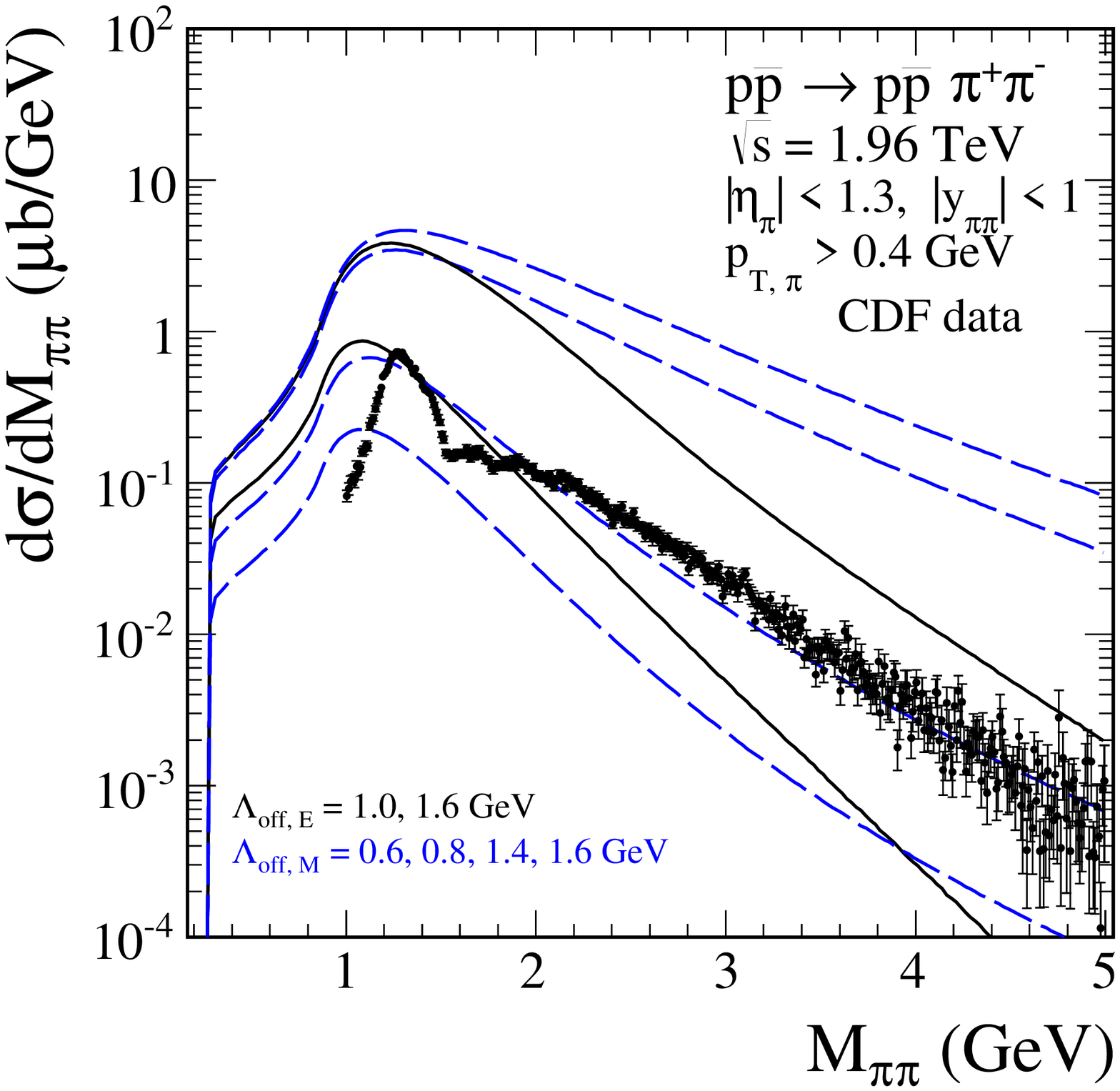}
\includegraphics[width=0.48\textwidth]{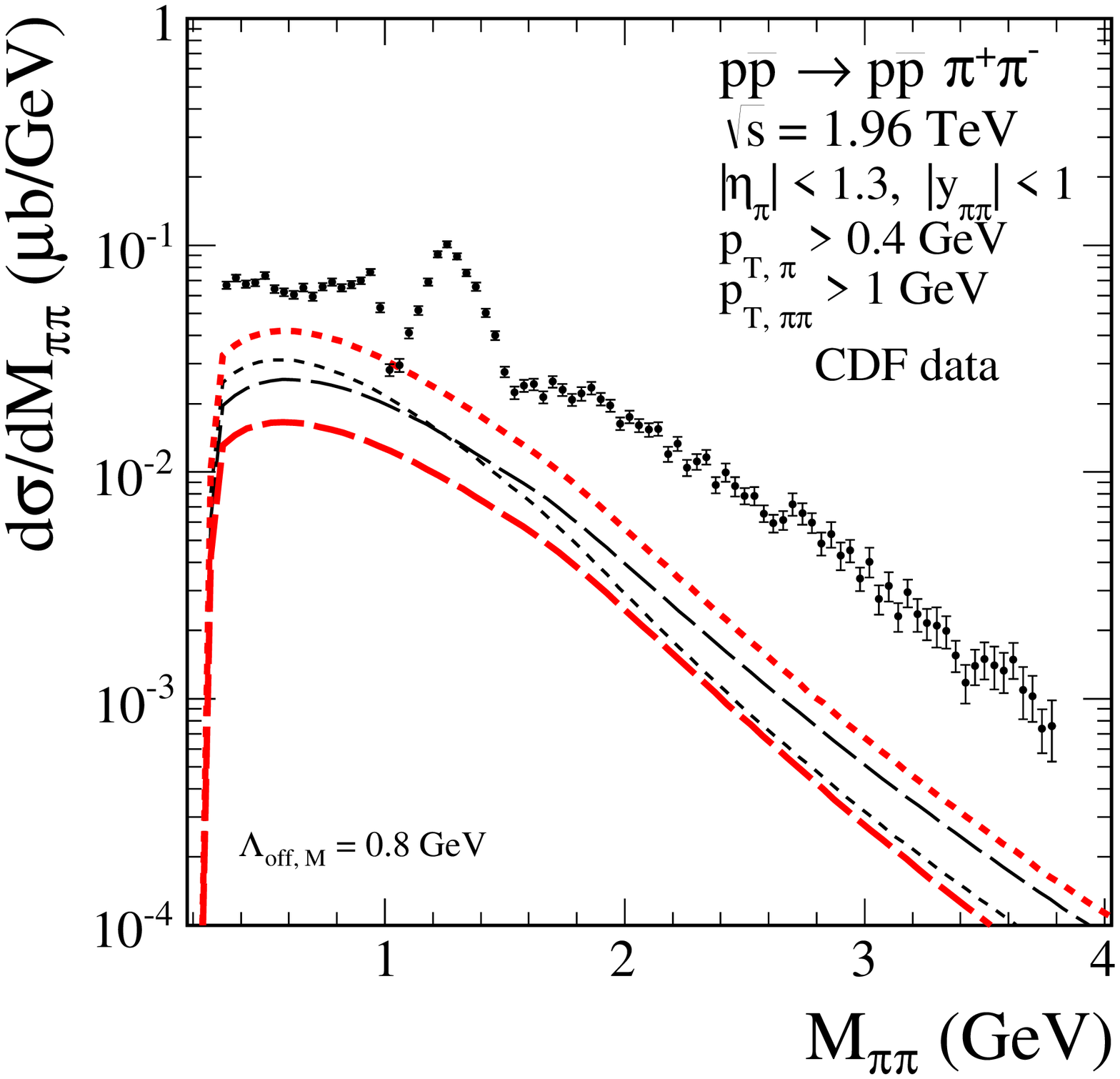}
  \caption{\label{fig:CDF_M34}
  \small
Two-pion invariant mass distribution at $\sqrt{s}=1.96$~TeV 
with the CDF kinematical cuts specified in the figure caption.
The meaning of the lines in the left panel is the same as in Fig.~\ref{fig:STAR_M34}.
In the right panel we show the results with an additional cut on transverse momentum of the pair $p_{t,\pi\pi} > 1$~GeV
without (dotted lines) and with (long-dashed lines) the $\pi N$ absorption corrections
and for two different $t$-dependences of the $\pi p$-subsystem amplitudes.
The black thin lines show results with formula (\ref{MN_slope_parameters})
while the red thick lines represent results for the replacement (\ref{modified_t_dependence_orear}) 
($B_{0} = 6.5$~GeV$^{-2}$, $\mu^{2} = 0.6$~GeV$^{2}$).
The CDF data \cite{Aaltonen:2015uva,Albrow_Project_new}
are shown with only statistical errors; systematic uncertainties are approximately 10\% at all masses.
}
\end{figure}

We wish to emphasize that in this experiment, in contrast to the STAR experiment, 
the final-state nucleons are not detected and only rapidity
gap conditions ($\Delta \eta > 4.6$ on each side of the $\pi^{+} \pi^{-}$) was imposed experimentally.
In Fig.~\ref{fig:CDF_M34} we show the two-pion invariant mass
distribution at $\sqrt{s}=1.96$~TeV for the $p \bar{p} \to p \bar{p} \pi^+ \pi^-$ reaction
and with the following CDF cuts on kinematical variables:
$p_{t, \pi} > 0.4$~GeV, $|\eta_{\pi}| < 1.3$ for both mesons, and $|y_{\pi\pi}| < 1$.
The rapidity of the central $\pi^{+}\pi^{-}$ system is expressed by the formula
\begin{eqnarray}
y_{\pi\pi} = \frac{1}{2} \ln \left( \frac{(p_{30}+p_{40})+(p_{3z}+p_{4z})}
                                         {(p_{30}+p_{40})-(p_{3z}+p_{4z})} \right) \,,
\end{eqnarray}             
with the four-momenta $p_{3}$ ($\pi^{+}$ meson) and $p_{4}$ ($\pi^{-}$ meson).
The kinematical cuts $p_{t, \pi} > 0.4$~GeV on both pions strongly 
distort the region of low $M_{\pi\pi} < 1$~GeV.
At $M_{\pi\pi} \simeq 1$~GeV the data show a minimum 
due to interference of the $f_{0}(980)$ resonance contribution
with the non-resonant background contribution.
At higher $M_{\pi\pi}$, in the region of 1.2-1.7~GeV, some structures
could be attributed to $f_{2}(1270)$, $f_{0}(1370)$, $f_{0}(1500)$, and $f_{0}(1710)$
resonant states.
The $f_{0}(1500)$ and the $f_{0}(1710)$ mesons 
are considered to be scalar glueball candidates \cite{Ochs:2013gi}, 
but mixing with quarkonium states complicating the issue.
We roughly describe the differential cross section
in the left panel when using the form factors 
(\ref{off-shell_form_factors_mon}) with $\Lambda_{off,M} \simeq 0.8$~GeV.
The data at $\sqrt{s} = 0.9$~TeV look similar (see Fig.~1 of \cite{Aaltonen:2015uva}).
The data at both energies include diffractive dissociation 
of proton and antiproton (all the produced unobserved hadrons have $|\eta|>5.9$),
so that low diffractive masses of the baryonic systems
are included, especially at $\sqrt{s} = 1.96$~TeV.
In the right panel of Fig.~\ref{fig:CDF_M34} we show results
with an extra cut on the $\pi^{+}\pi^{-}$ transverse momentum.
The results for the form factors that give a reasonable description
in the left panel badly fail to describe the data in the right panel,
underestimating the CDF data by a factor of about 5.
In this case our model results are much below the experimental data
which could be due to a contamination of non-exclusive processes
or the perturbative mechanism discussed in \cite{HarlandLang:2012qz}.
Both the interference of resonant state with the $\pi^{+}\pi^{-}$-continuum and 
the diffractive dissociation effects require more subtle theoretical approach.
This will be addressed elsewhere.
Thus, the non-resonant Lebiedowicz-Szczurek model should not be expected to fit the data precisely.
We conclude that the CDF data for $p_{t,\pi \pi} > 1$~GeV \cite{Aaltonen:2015uva} 
are sensitive to the details of the ``large''-$t$ behavior of the $\pi N$ scattering. 
While neglecting the extra $\pi N$ absorptive corrections 
the stretched exponential functional form gives larger
cross section than the standard exponential form,
however, the inclusion of the extra $\pi N$ absorption cancels the ``improvement''.
In this moment we do not know a solution of this puzzle.

\begin{figure}[!ht]
\includegraphics[width=0.49\textwidth]{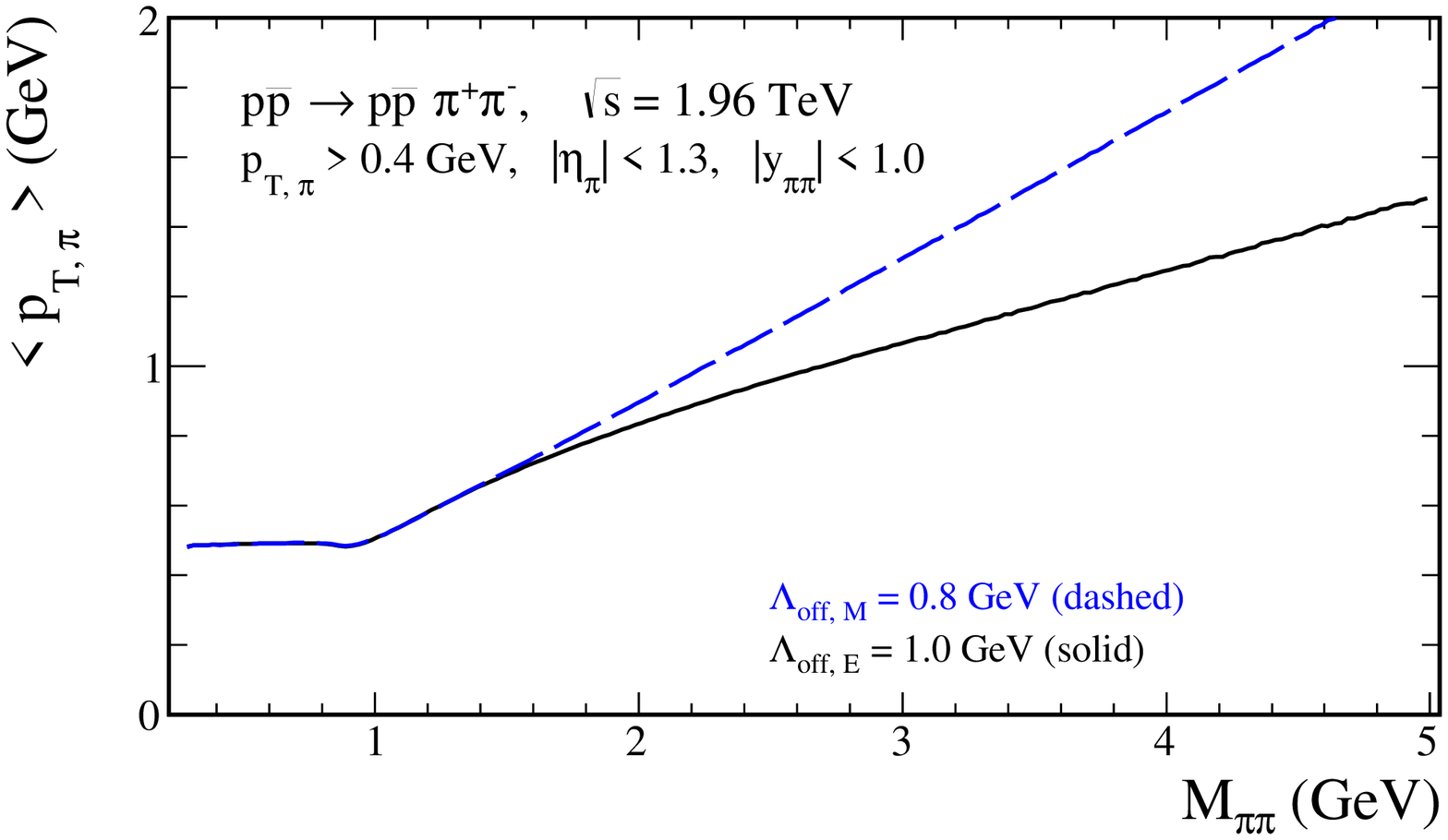}
\includegraphics[width=0.49\textwidth]{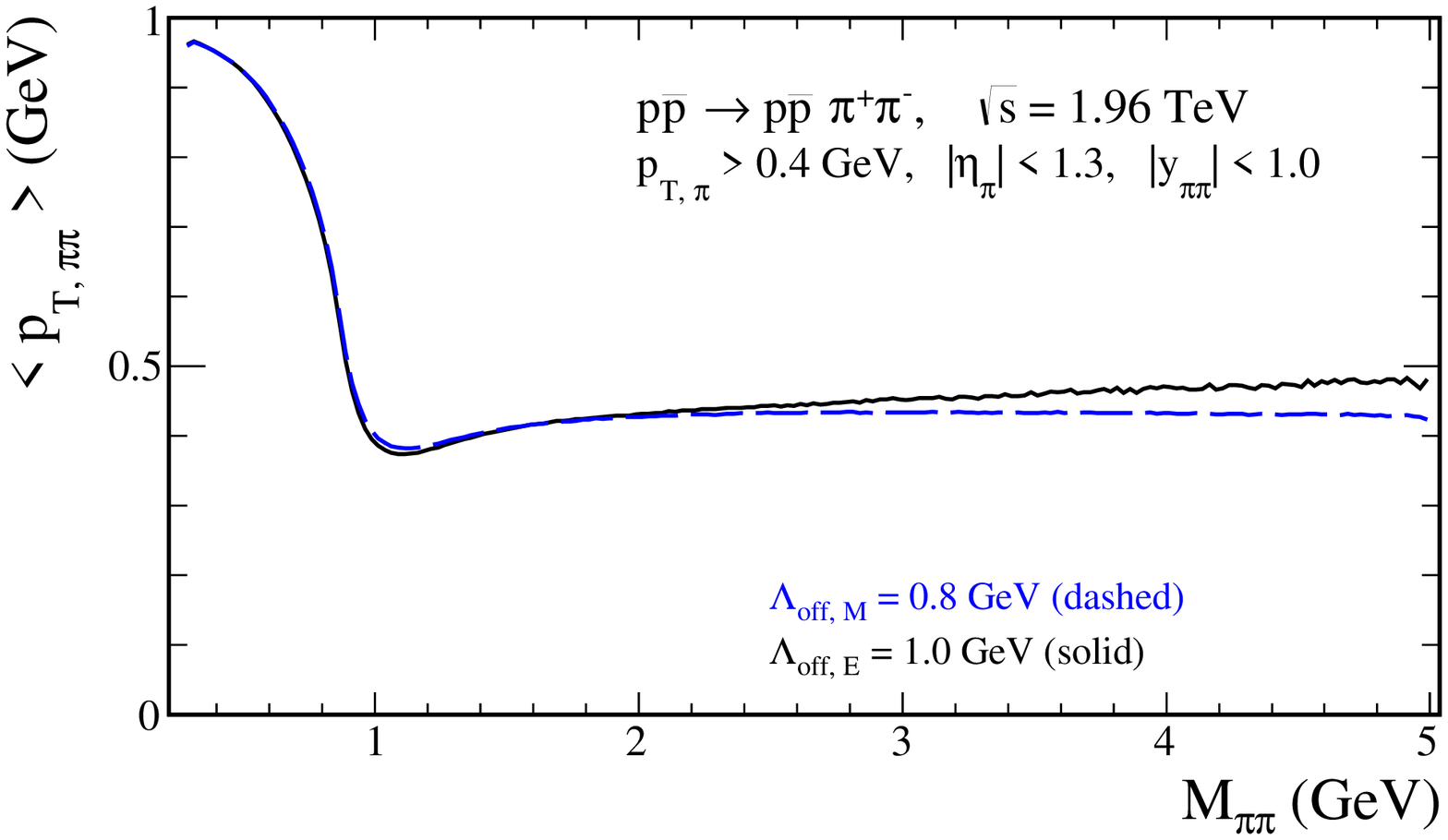}
  \caption{\label{fig:CDF_ave_pt3_ave_ptsum}
  \small
Mean value of $p_{t, \pi}$ (left panel) 
and $p_{t, \pi\pi}$ (right panel) as a function of two-pion 
invariant mass at $\sqrt{s} = 1.96$~TeV calculated
with the CDF kinematical cuts specified in the figure caption.
}
\end{figure}

Now let us discussed shortly quantities or observables that are
sensitive to the pion off-shell form factors.
The dependence of $<p_{t, \pi}>$ and $<p_{t, \pi\pi}>$ 
as a function of two-pion invariant mass is presented in 
Fig.~\ref{fig:CDF_ave_pt3_ave_ptsum}.
Our calculation shows a rise of the average pion transverse momentum with
dipion invariant mass. A dependence on the form of the form factor
is clearly seen. On the contrary, the average transverse momentum of the
dipion pair is almost independent of the form of the form factor and a parameter
of the form factor. This can be understood from momentum conservation.
The transverse momentum of the dipion system must be balanced by
the transverse momenta of protons. The latter distributions (shapes) are
obviously independent of the pion off-shell form factors.

\begin{figure}[!ht]  
\center
\includegraphics[width=0.49\textwidth]{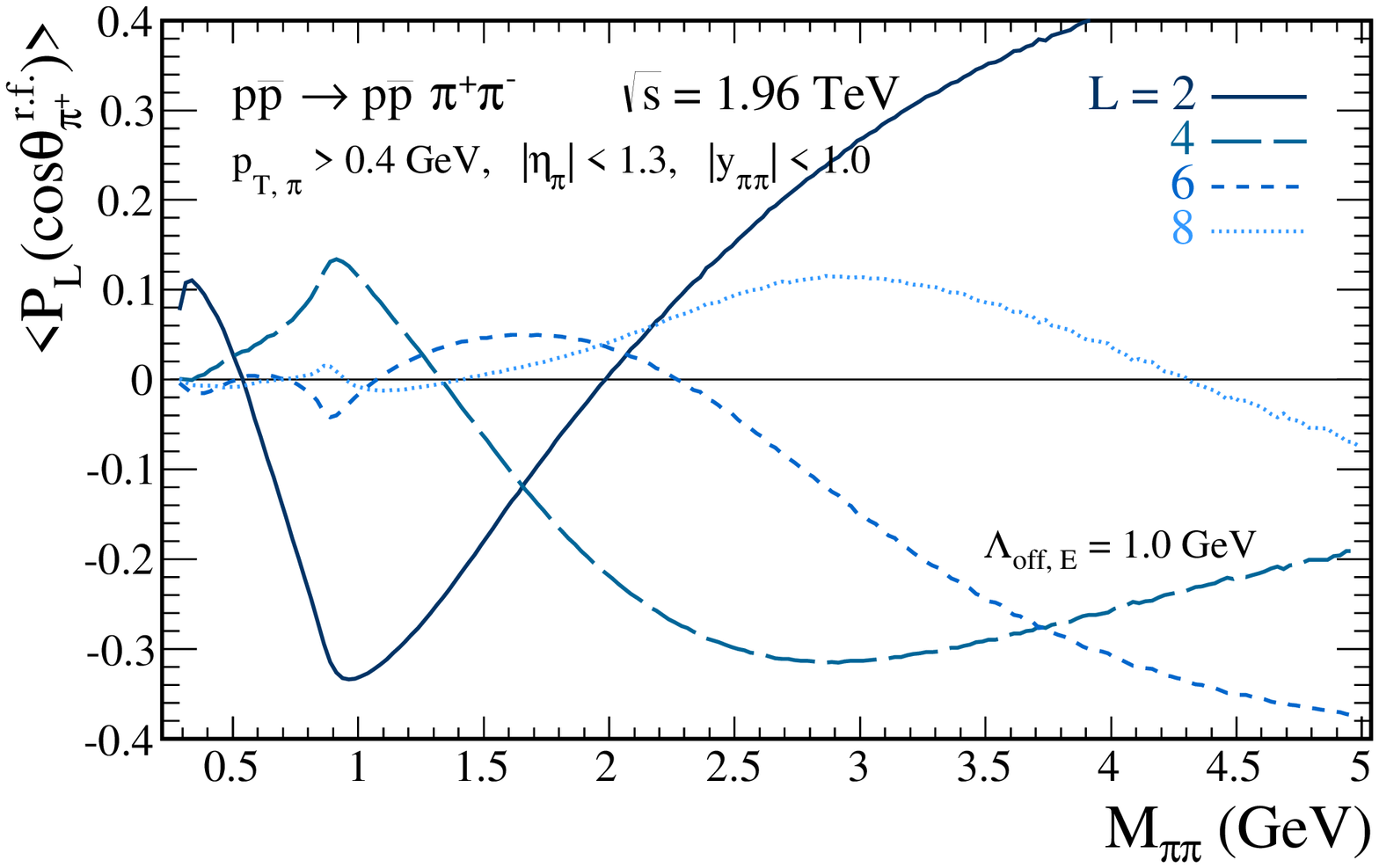}
\includegraphics[width=0.49\textwidth]{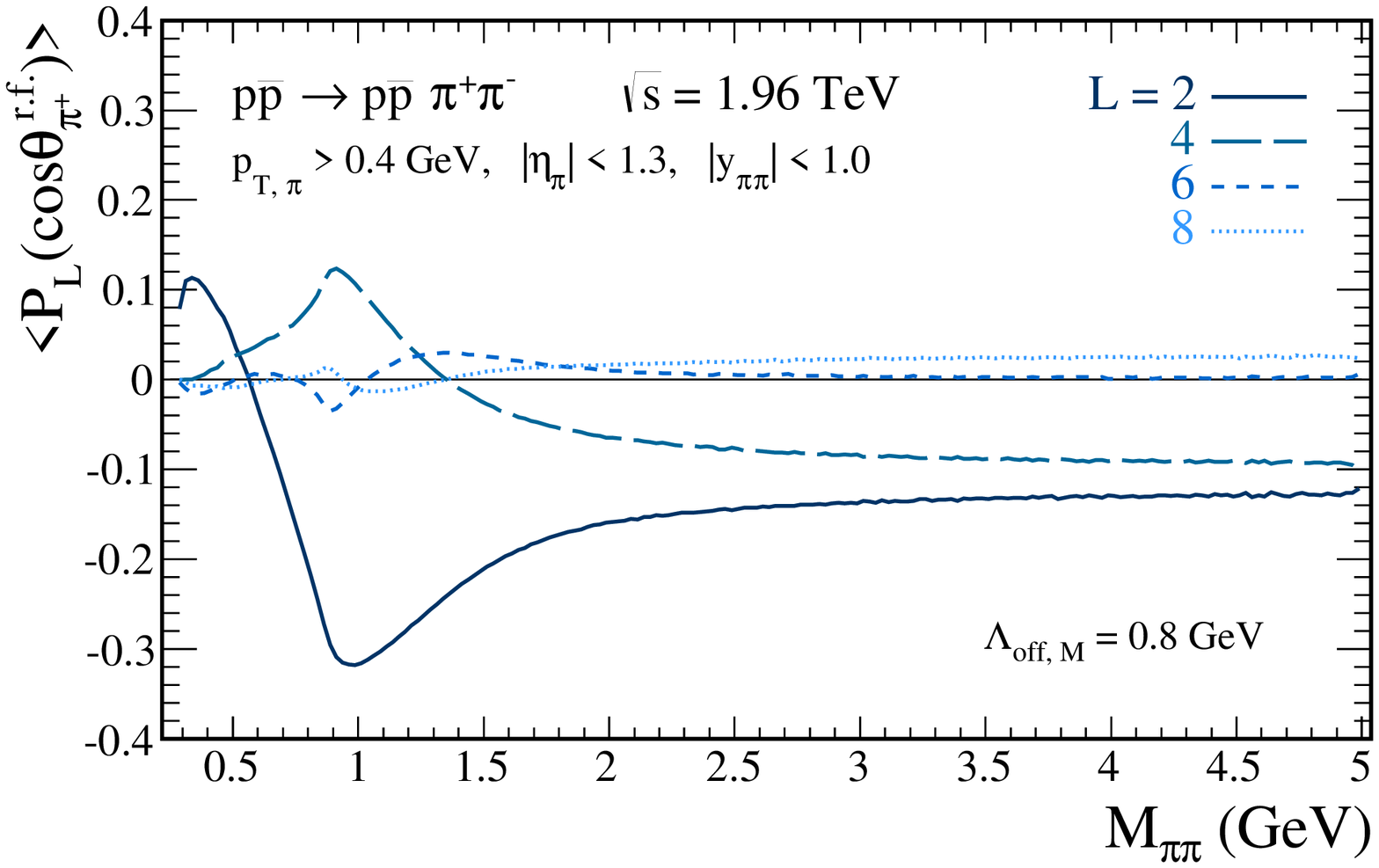}
  \caption{\label{fig:CDF_PevenL_M34}
  \small
Mean value of the first even Legendre polynomials 
$P_{L}(\cos\theta_{\pi^{+}}^{\,r.f.})$
as a function of two-pion invariant mass with the CDF kinematical cuts 
specified in the figure caption.
The results correspond to two types of off-shell pion form factors:
the exponential one (\ref{off-shell_form_factors_exp}) 
and the monopole one (\ref{off-shell_form_factors_mon}).
}
\end{figure}

Another observable which can be very sensitive 
to the choice of off-shell pion form factors are the Legendre polynomials
$<P_{L_{even}}(\cos\theta_{\pi^{+}}^{ r.f.})>(M_{\pi\pi})$ distributions,
where $cos\theta_{\pi^{+}}^{ \,r.f.}$ is the angle of the $\pi^{+}$ meson 
with respect to the beam axis, in the $\pi^{+}\pi^{-}$ rest frame.
In Fig.~\ref{fig:CDF_PevenL_M34} we present
the average $P_{L}$ calculated as
\begin{eqnarray}
<P_{L}(\cos\theta_{\pi^{+}}^{\,r.f.})>(M_{\pi\pi}) = 
\frac{\int d\mathcal{PS} \; P_{L}(\cos\theta_{\pi^{+}}^{\,r.f.}) \; d\sigma/d\mathcal{PS}(M_{\pi\pi})}
     {\int d\mathcal{PS} \; d\sigma/d\mathcal{PS}(M_{\pi\pi})} \,,
\end{eqnarray}             
where the integral is done over experimental phase space.
We have found that the $<P_{L}(\cos\theta^{\,r.f.})>(M_{\pi\pi})$ 
distributions are almost unaffected by the absorption effects.
%
%
The difference between the results for form factors 
(\ref{off-shell_form_factors_exp}) and (\ref{off-shell_form_factors_mon})
is huge at higher invariant masses and thus such observables 
may prove very useful in distinguishing between these choices.
Experimental results of the $<P_{L}(\cos\theta_{\pi^{+}}^{\,r.f.})>(M_{\pi\pi})$ 
distributions for $\sqrt{s} = 1.96$~TeV
are presented in Fig.~23 of \cite{Albrow_Project_new}
and strongly support our predictions calculated 
with the monopole form factors (\ref{off-shell_form_factors_mon}) 
and the cut-off parameter $\Lambda_{off,M} = 0.8$~GeV,
particularly at higher two-pion invariant masses $M_{\pi\pi} > 1.5$~GeV;
see also discussion in section~2.6.2 (Tevatron) of \cite{Lebiedowicz:thesis}.
One can observe in Fig.~2.42 of \cite{Lebiedowicz:thesis} 
that the contribution of $L = 4$ is small
at low $M_{\pi\pi}$ when the cuts are neglected (left panels) and significant 
already at $M_{\pi\pi} \approx 1$~GeV when the cuts are applied (right panels).
This suggests that the CDF kinematic cuts may distort the partial wave content.
This makes conclusions more difficult.

\subsection{ALICE experiment}

\begin{figure}[!ht]
\includegraphics[width=0.48\textwidth]{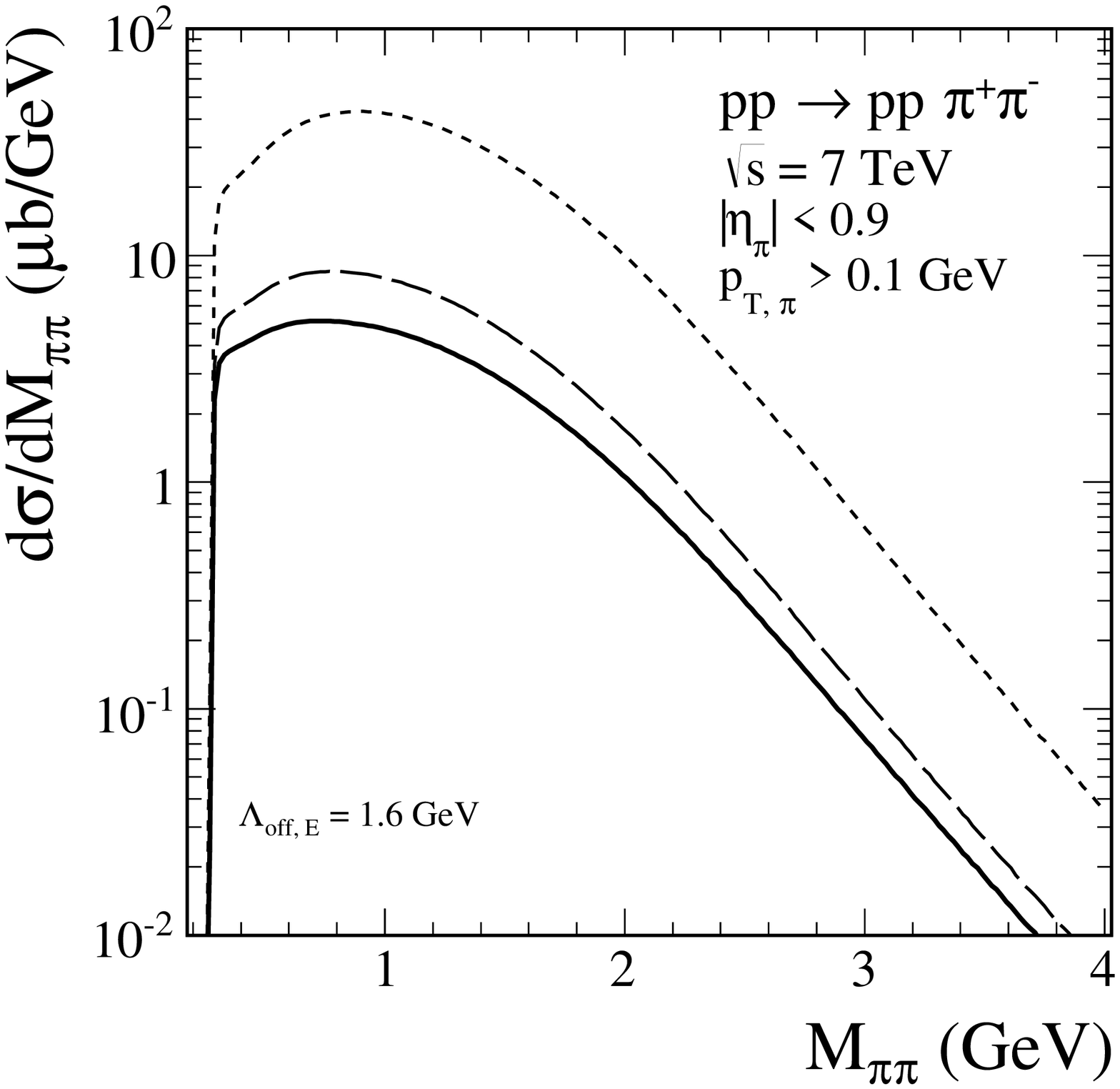}
\includegraphics[width=0.48\textwidth]{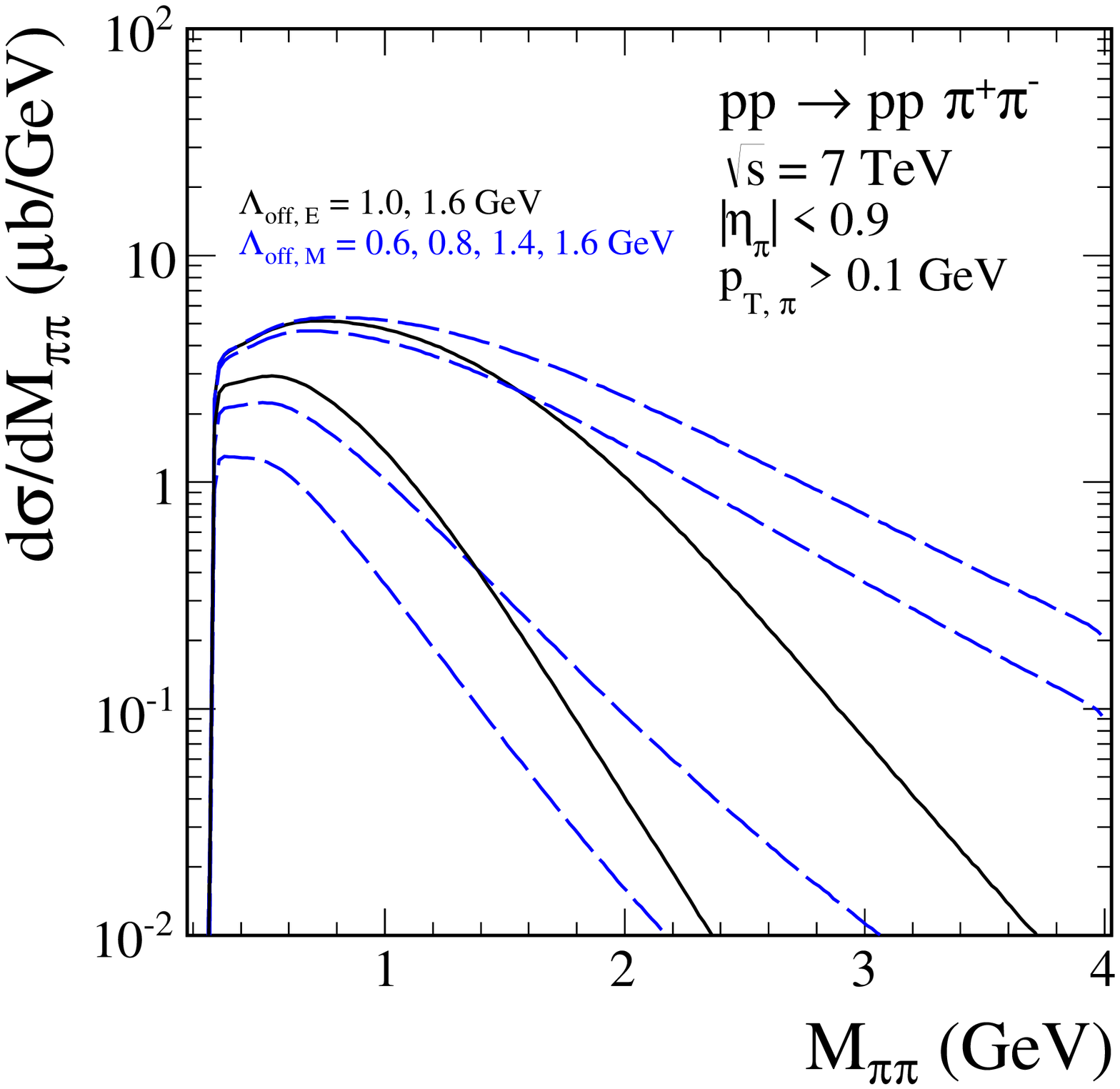}
  \caption{\label{fig:ALICE_M34}
  \small
Two-pion invariant mass distribution at $\sqrt{s}=7$~TeV 
with the ALICE kinematical cuts specified in the figure caption.
In the left panel we show results 
with a cut-off parameter 
$\Lambda_{off,E} = 1.6$~GeV, see (\ref{off-shell_form_factors_exp}),
without the absorption effects (the dotted line),
with the $pp$-rescattering (the long-dashed line),
and with the additional $\pi p$-rescattering (the solid line).
The meaning of the lines in the right panel is the same as in Fig.~\ref{fig:STAR_M34}.
}
\end{figure}

Now, we shall present our predictions for experiments at the LHC.
We shall start review of our results for the case of the ALICE experiment at $\sqrt{s} = 7$~TeV.
We impose the corresponding cuts on both pions transverse momenta
$p_{t, \pi} > 0.1$~GeV and pseudorapidities $|\eta_{\pi}| < 0.9$. 
In Fig.~\ref{fig:ALICE_M34} we show two-pion invariant mass
distribution. As for the case of the STAR experiment in the left panel
we show the Born result (dotted line), the result with $pp$ absorption 
only (dashed line) as well as the results when including 
the extra $\pi p$ absorption (solid line). 
There is similar tendency as for the STAR case.
The extra absorption lower the cross section without modifying the
shape of the invariant mass distribution.
In the right panel we show our result for two different forms of
the off-shell form factor and different values of the cut-off
parameters. As for the STAR case the shape strongly depends
on the form factor form as well as on the cut-off parameters.

\begin{figure}[!ht]
\includegraphics[width=0.48\textwidth]{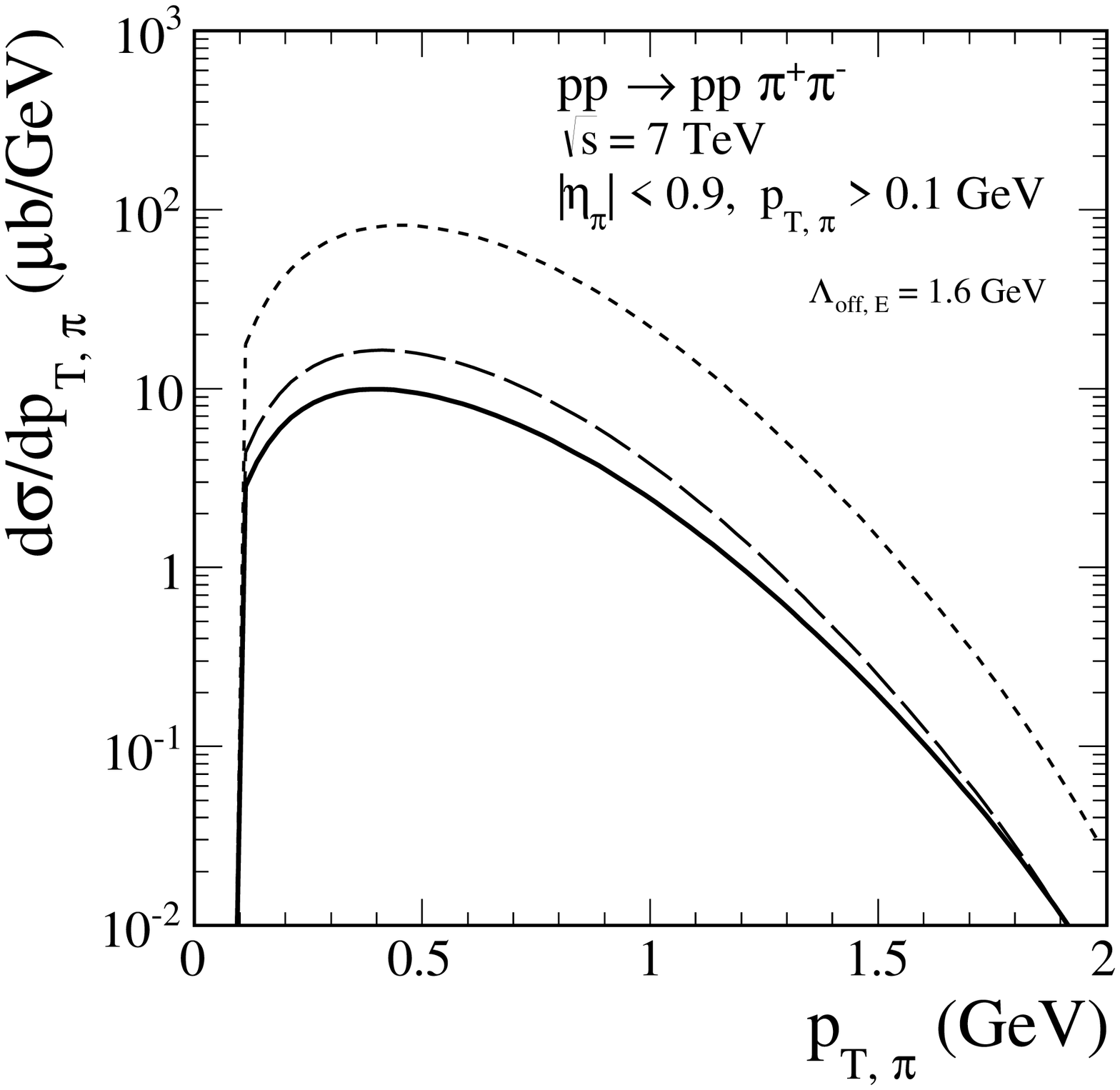}
\includegraphics[width=0.48\textwidth]{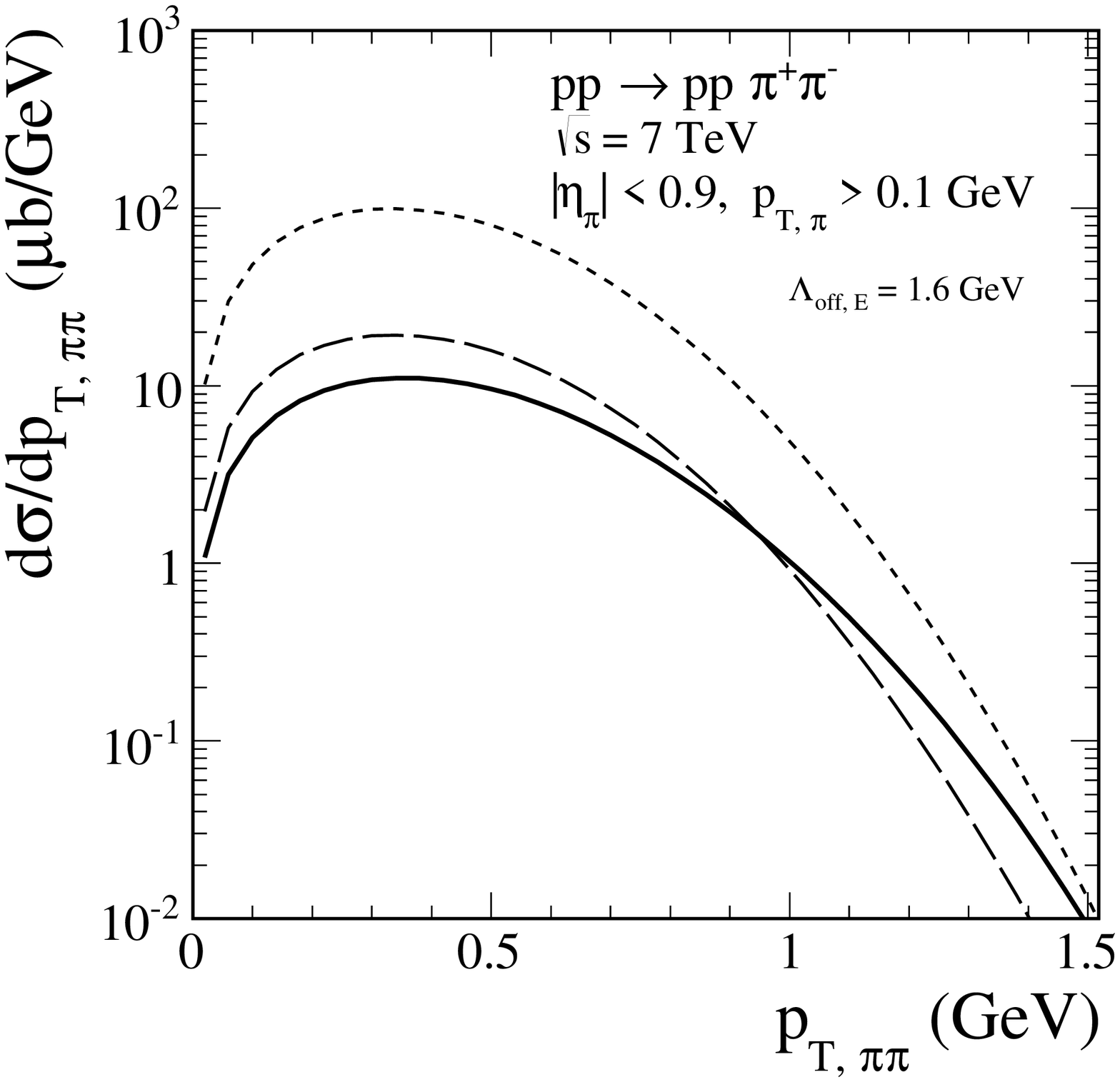}
\includegraphics[width=0.48\textwidth]{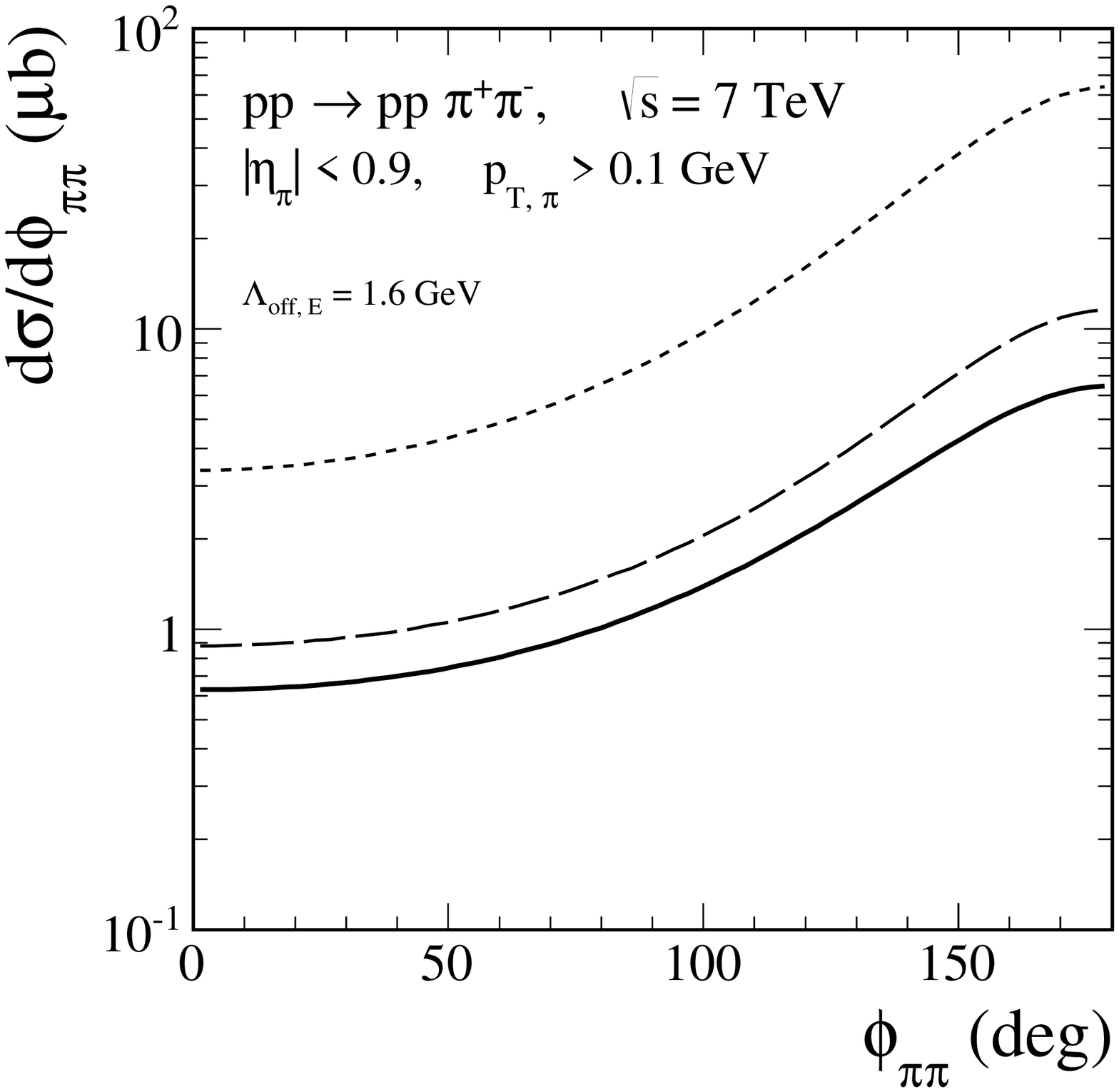}
  \caption{\label{fig:ALICE_phi34}
  \small
Differential cross sections 
$d\sigma/dp_{t,\pi}$, 
$d\sigma/dp_{t,\pi \pi}$, and $d\sigma/d\phi_{\pi \pi}$ at $\sqrt{s} = 7$~TeV 
with the ALICE kinematical cuts specified in the figure caption.
In the calculation we have used the cut-off parameter $\Lambda_{off,E} = 1.6$~GeV.
The meaning of the lines is the same as in Fig.~\ref{fig:STAR_M34} (left panel).
}
\end{figure}

Now we pass to distributions in transverse momenta of single pion and of the pion pair,
see Fig.~\ref{fig:ALICE_phi34} (top panels).
The absorption effects due to $\pi p$ interaction
change the shape of the $p_{t,\pi \pi}$-distribution.
Such a distribution can be easily measured by the ALICE collaboration. 
The ALICE experiment cannot register forward/backward protons.
Therefore only azimuthal correlations between pions can be measured.
Our corresponding distribution is shown in Fig.\ref{fig:ALICE_phi34} (bottom panel).
The $\phi_{\pi\pi}$ distribution peaks in the back-to-back
configuration, i.e. when $\phi_{\pi\pi} = \pi$.
The absorption effects practically do not change the shape of the distributions.

\begin{figure}[!ht]
\includegraphics[width=0.49\textwidth]{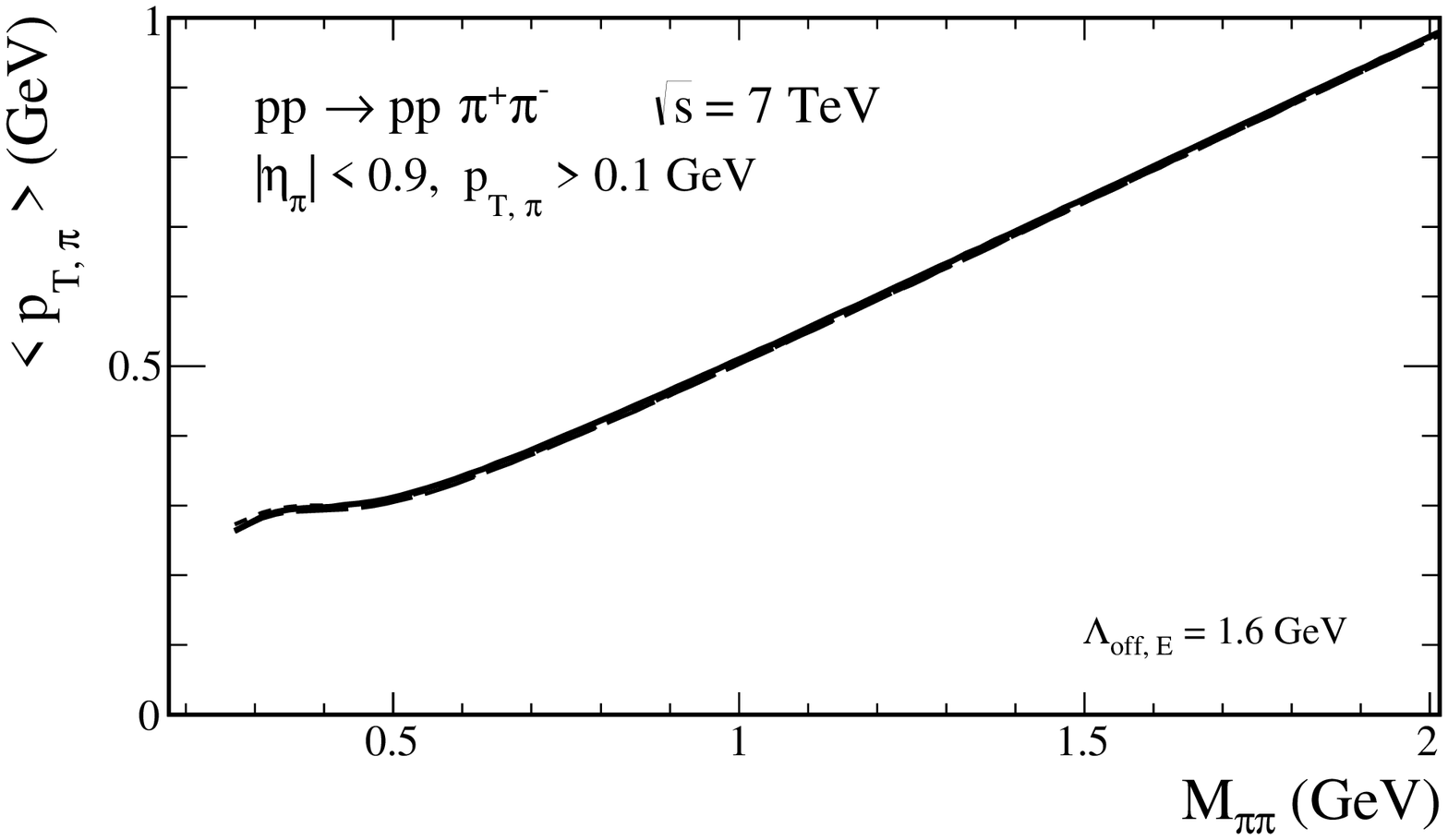}
\includegraphics[width=0.49\textwidth]{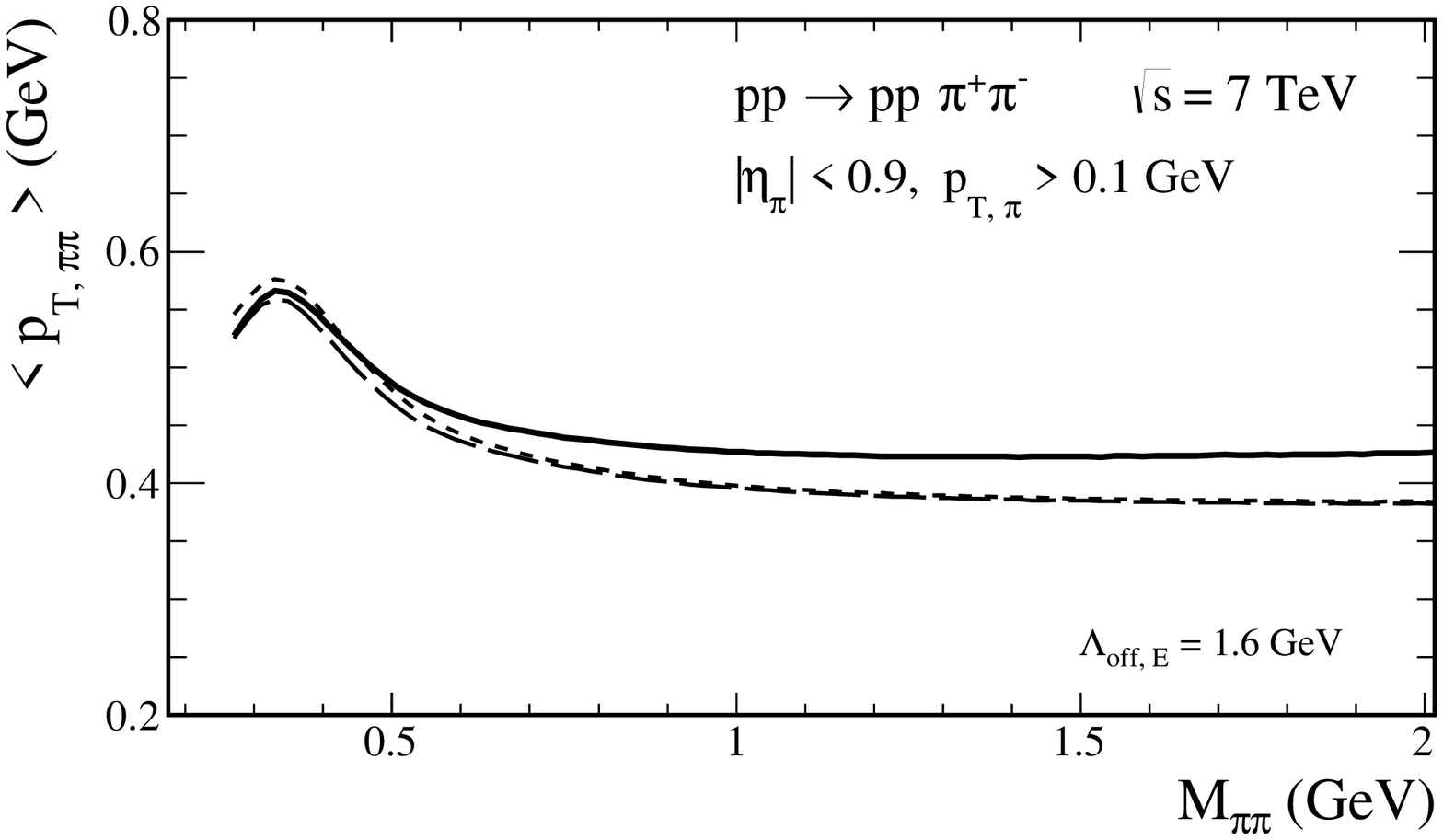}
  \caption{\label{fig:ALICE_ave_pt3_ave_ptsum}
  \small
Mean value of the $p_{t, \pi}$ (left panel) and $p_{t, \pi\pi}$ (right panel) 
as a function of two-pion invariant mass at $\sqrt{s} = 7$~TeV 
with the ALICE kinematical cuts specified in the figure caption.
In the calculation we have used the cut-off parameter 
$\Lambda_{off,E} = 1.6$~GeV.
The meaning of the lines is the same as in Fig.~\ref{fig:ALICE_phi34}.
}
\end{figure}

The average values of transverse momenta of single
pion $<p_{t, \pi}>(M_{\pi\pi})$ and of the pion pair $<p_{t, \pi\pi}>(M_{\pi\pi})$ 
are shown in Fig.~\ref{fig:ALICE_ave_pt3_ave_ptsum}. 
The results have been obtained assuming that $p_{t,\pi} >0.1$~GeV
without the absorption effects (the dotted line),
with the $pp$-rescattering (the long-dashed line),
and with the additional $\pi p$-rescattering (the solid line).

\subsection{CMS and ATLAS experiments}

The ATLAS tracking detector provides measurement of
charged particle momenta in the $|\eta| < 2.5$ region. 
Since the correlation between the pseudorapidities of both pions is very large, 
the measurement can be performed independently using the tracking detector ($|\eta| < 2.5$) 
and the forward calorimeters ($2.5 < |\eta| < 4.9$), see Fig.~4 of \cite{Staszewski:2011bg}.
We wish to note that the analysis in \cite{Staszewski:2011bg} was performed for 
$\Lambda_{off,E}^{2} = 2$~GeV$^{2}$ neglecting effect of the $\pi N$ rescattering.
Below we shall show results of non-resonant model (including all rescattering corrections)
for the CMS experiment and the corresponding kinematics cuts on both pions:
$p_{t, \pi} > 0.1$~GeV and $|\eta_{\pi}| < 2.0$. 
The general features of the differential distributions 
for the ATLAS experiment are, however, similar.

\begin{figure}[!ht]
\includegraphics[width=0.48\textwidth]{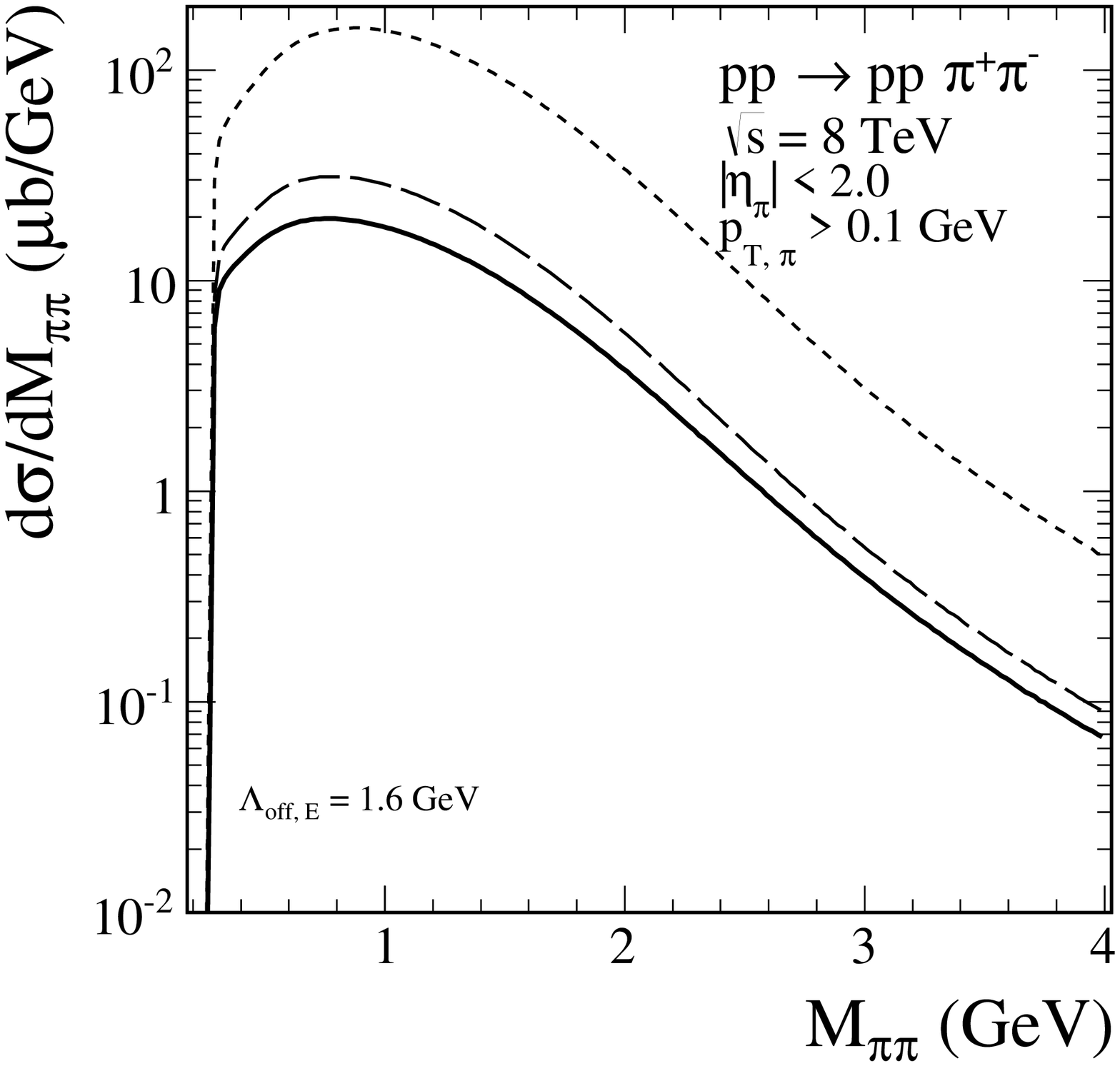}
\includegraphics[width=0.48\textwidth]{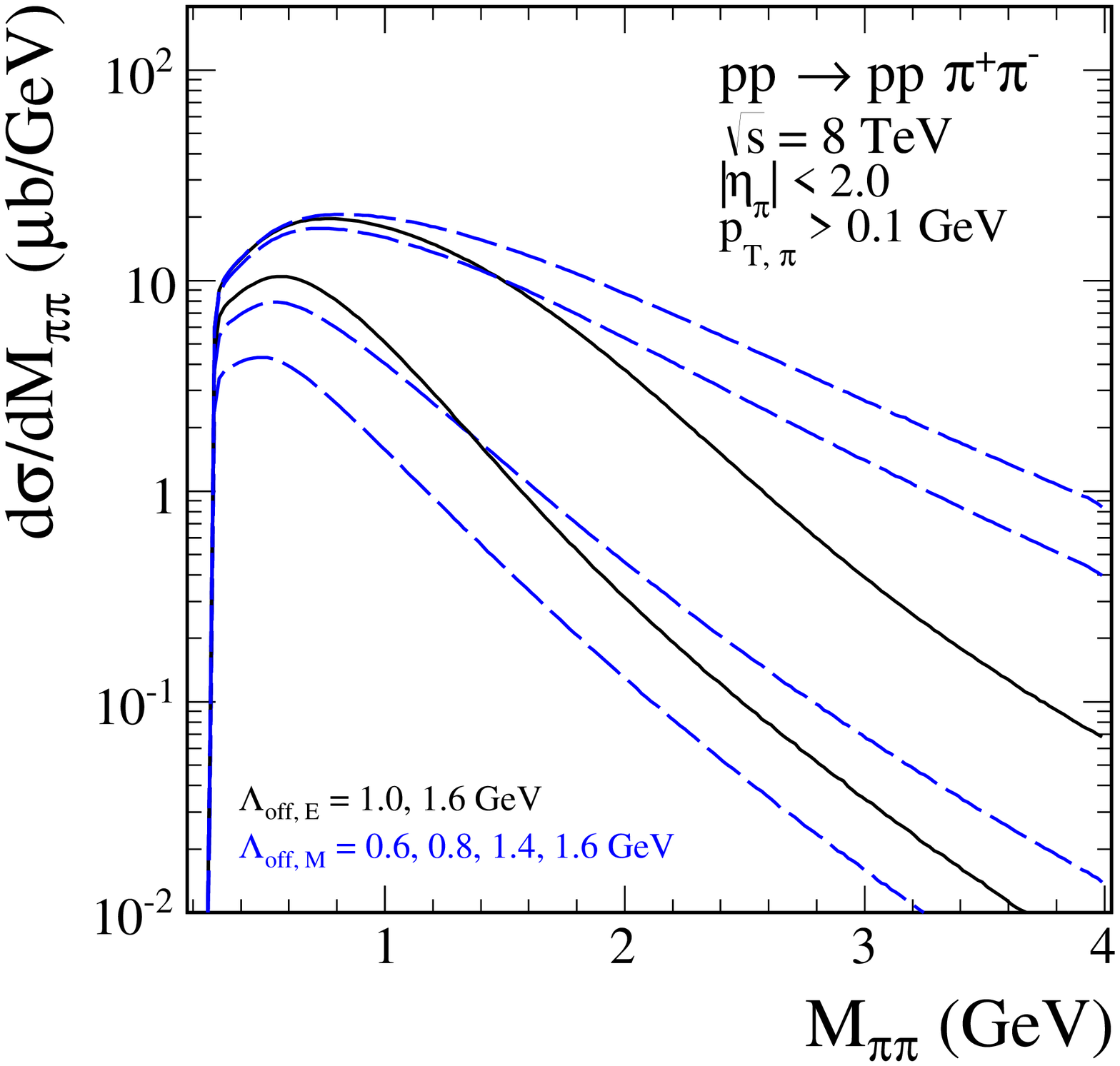}
  \caption{\label{fig:CMS_M34}
  \small
Two-pion invariant mass distribution at $\sqrt{s}=8$~TeV 
with the CMS kinematical cuts specified in the figure caption.
The meaning of the lines is the same as in Fig.~\ref{fig:STAR_M34}.
}
\end{figure}

In Fig.~\ref{fig:CMS_M34} we show two-pion invariant mass distribution.
In the left panel we show again results for three cases: 
Born (dashed line), absorption due to $pp$ interaction (dashed line) 
and for the case with extra $\pi p$ interaction (solid line). 
In the right panel we show the dependence of the cross section
on the choice of the pion off-shell form factor.

\begin{figure}[!ht]
\includegraphics[width=0.43\textwidth]{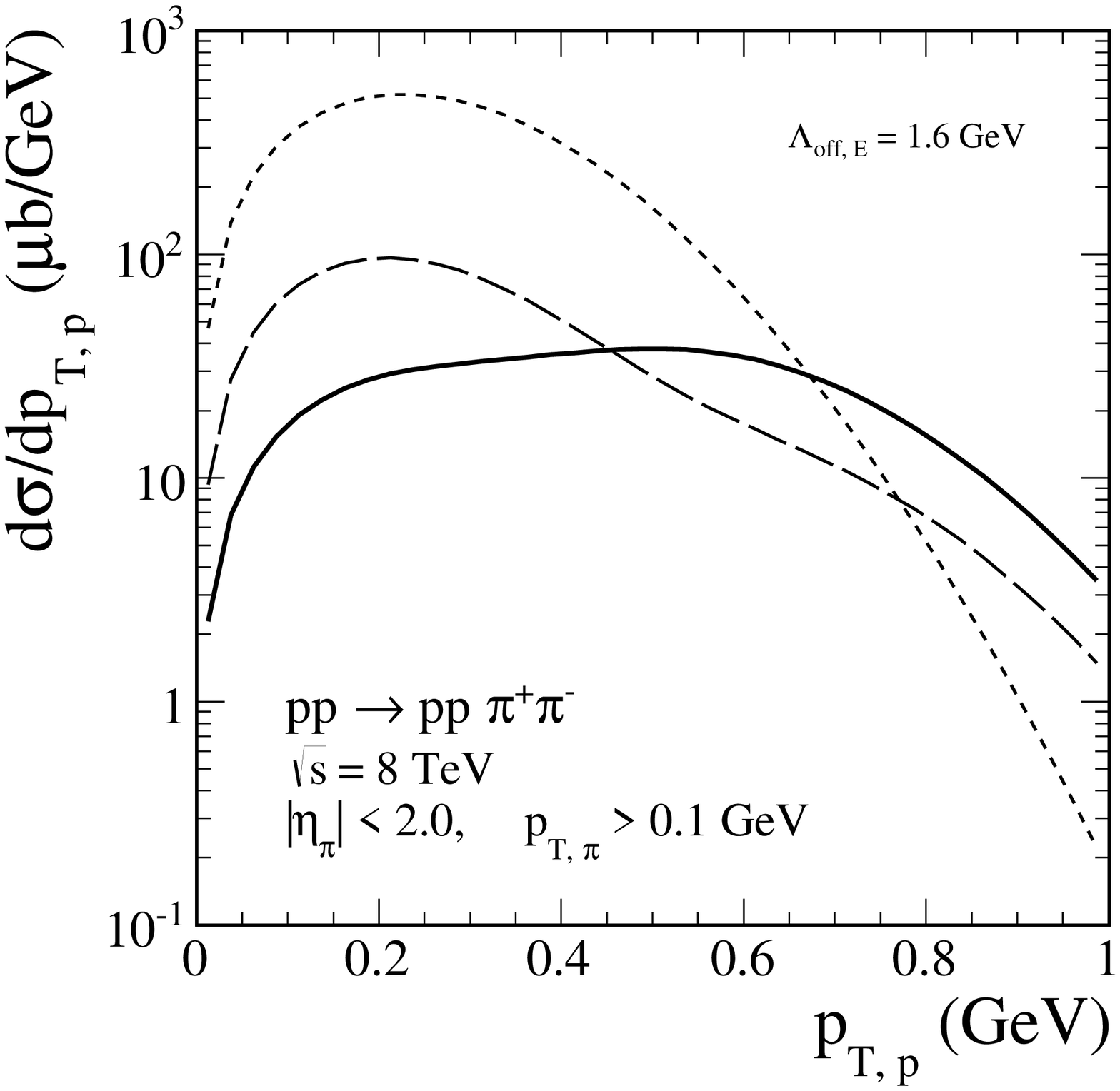}
\includegraphics[width=0.43\textwidth]{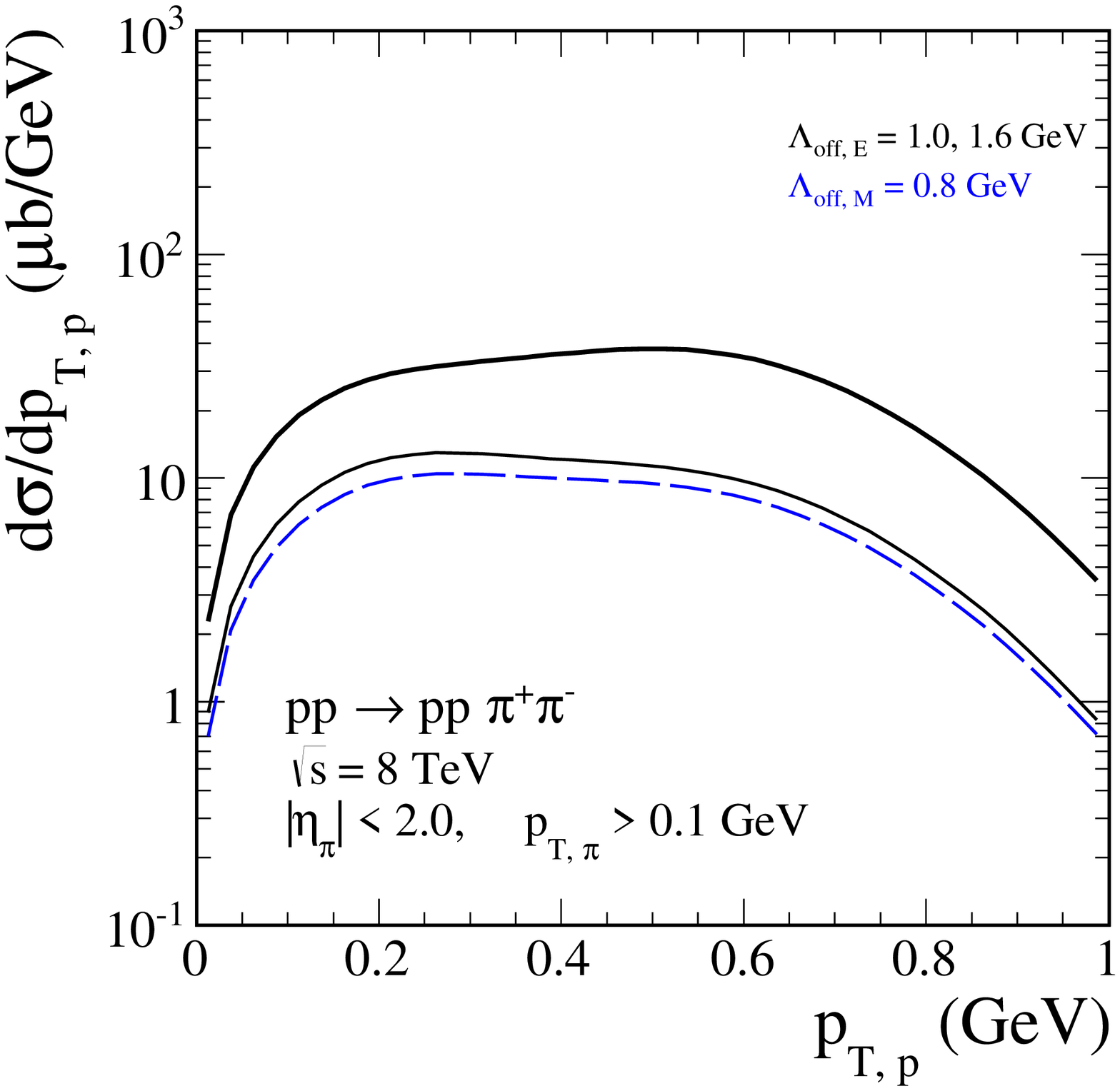}\\
\includegraphics[width=0.43\textwidth]{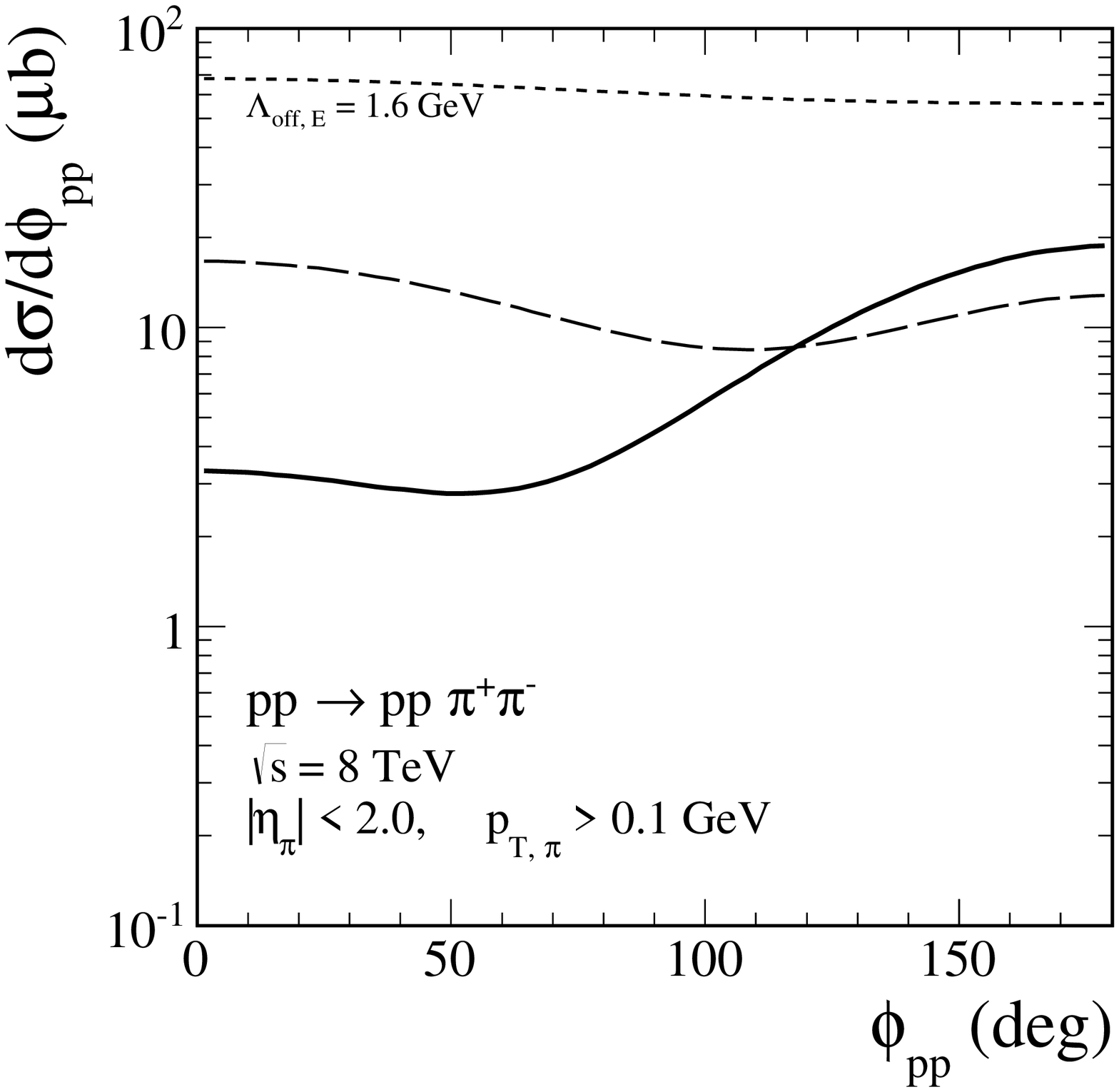}
\includegraphics[width=0.43\textwidth]{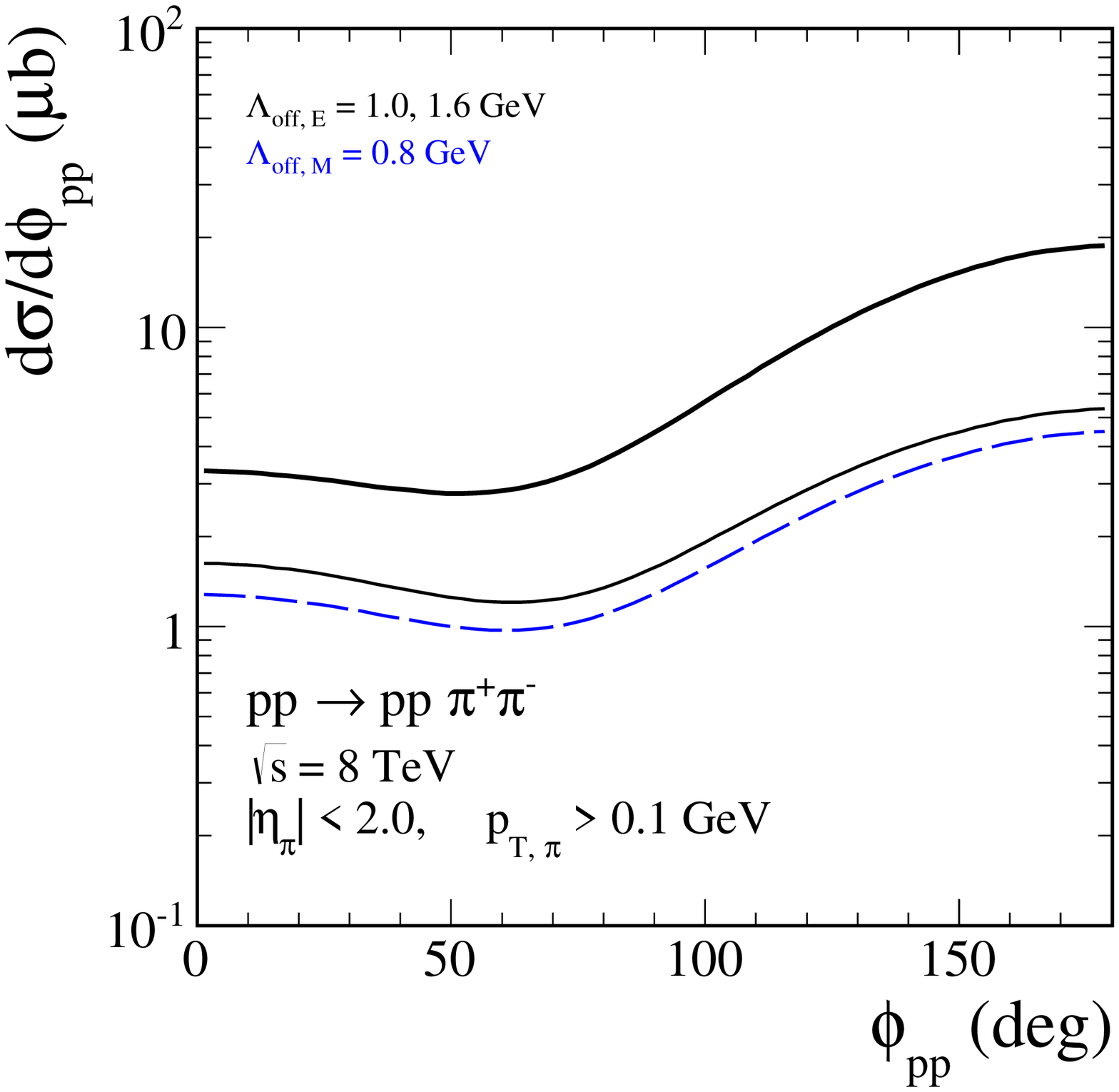}\\
\includegraphics[width=0.43\textwidth]{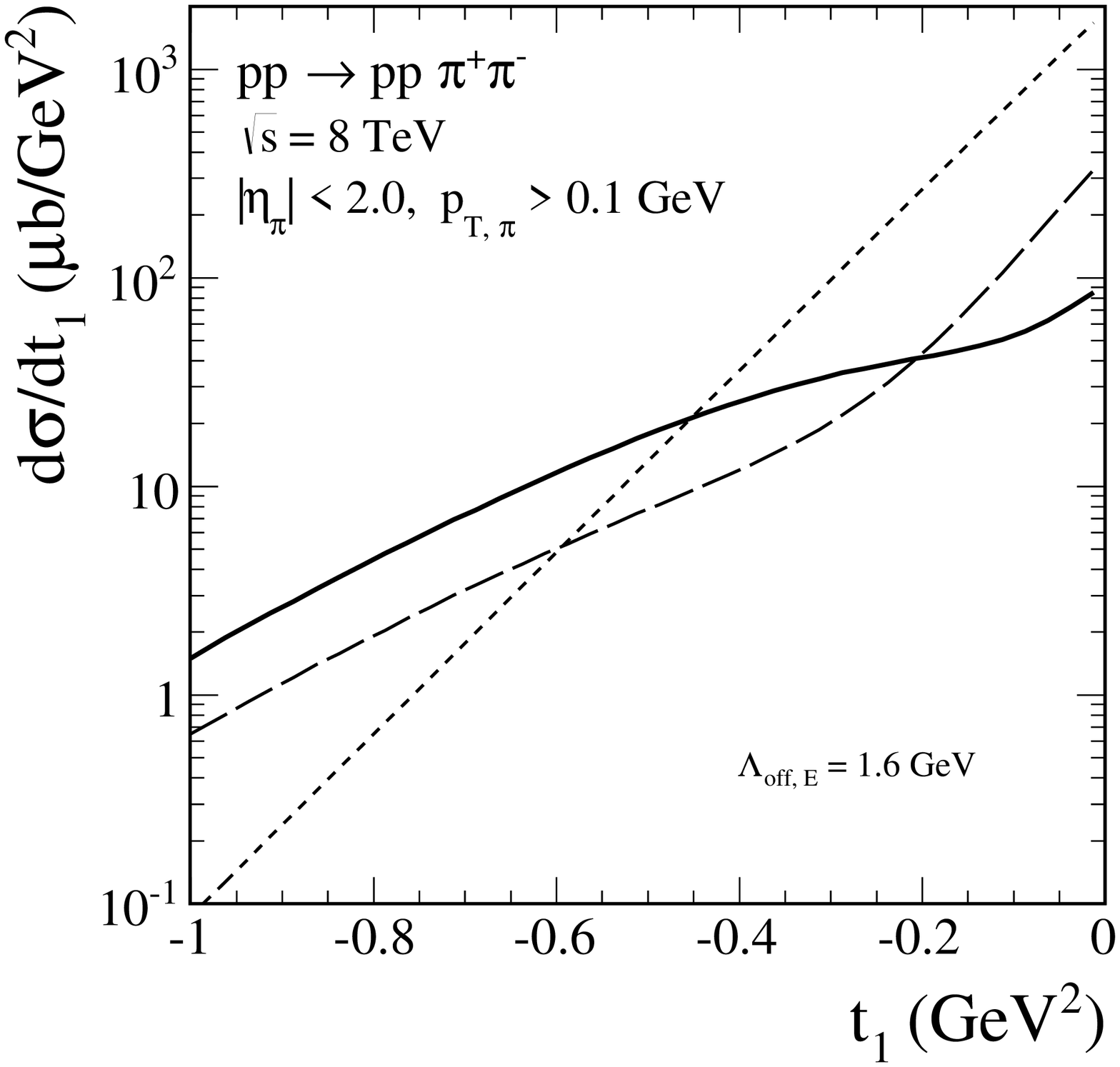}
\includegraphics[width=0.43\textwidth]{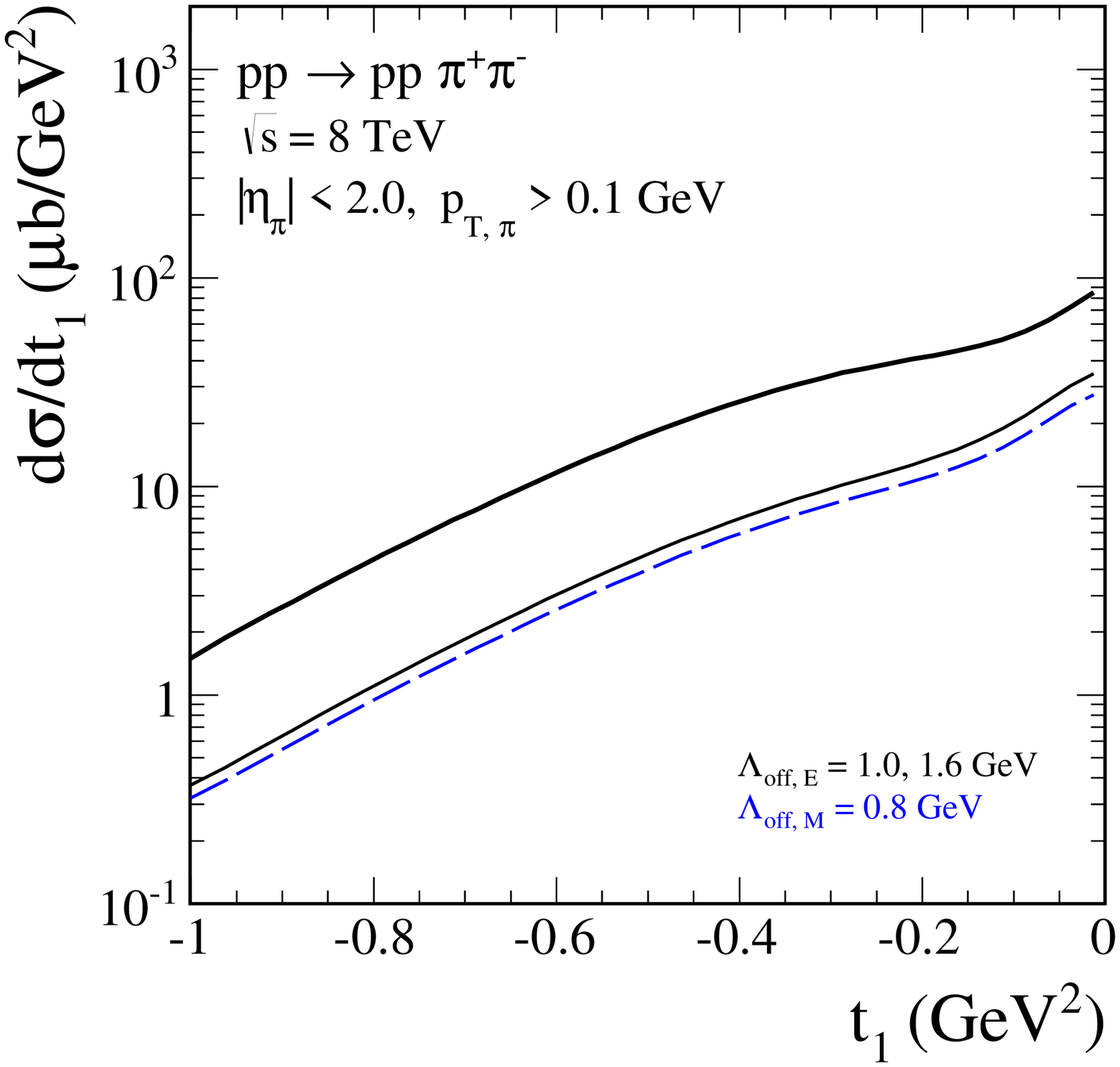}
  \caption{\label{fig:CMS_pt1_phi12}
  \small
The distributions in proton transverse momentum (top panels),
in azimuthal angle between the outgoing protons (middle panels),
and in proton four-momentum transfer $t_{1}$ (bottom panels) at $\sqrt{s} = 8$~TeV 
with the CMS kinematical cuts specified in the figure caption.
In the left panels we show the distributions without
and with the absorption corrections.
In the calculation results of which are shown on the right panel we have used 
two form for the off-shell pion form factors and different cut-off parameters $\Lambda_{off}$.
}
\end{figure}

\begin{figure}[!ht]
\includegraphics[width=0.45\textwidth]{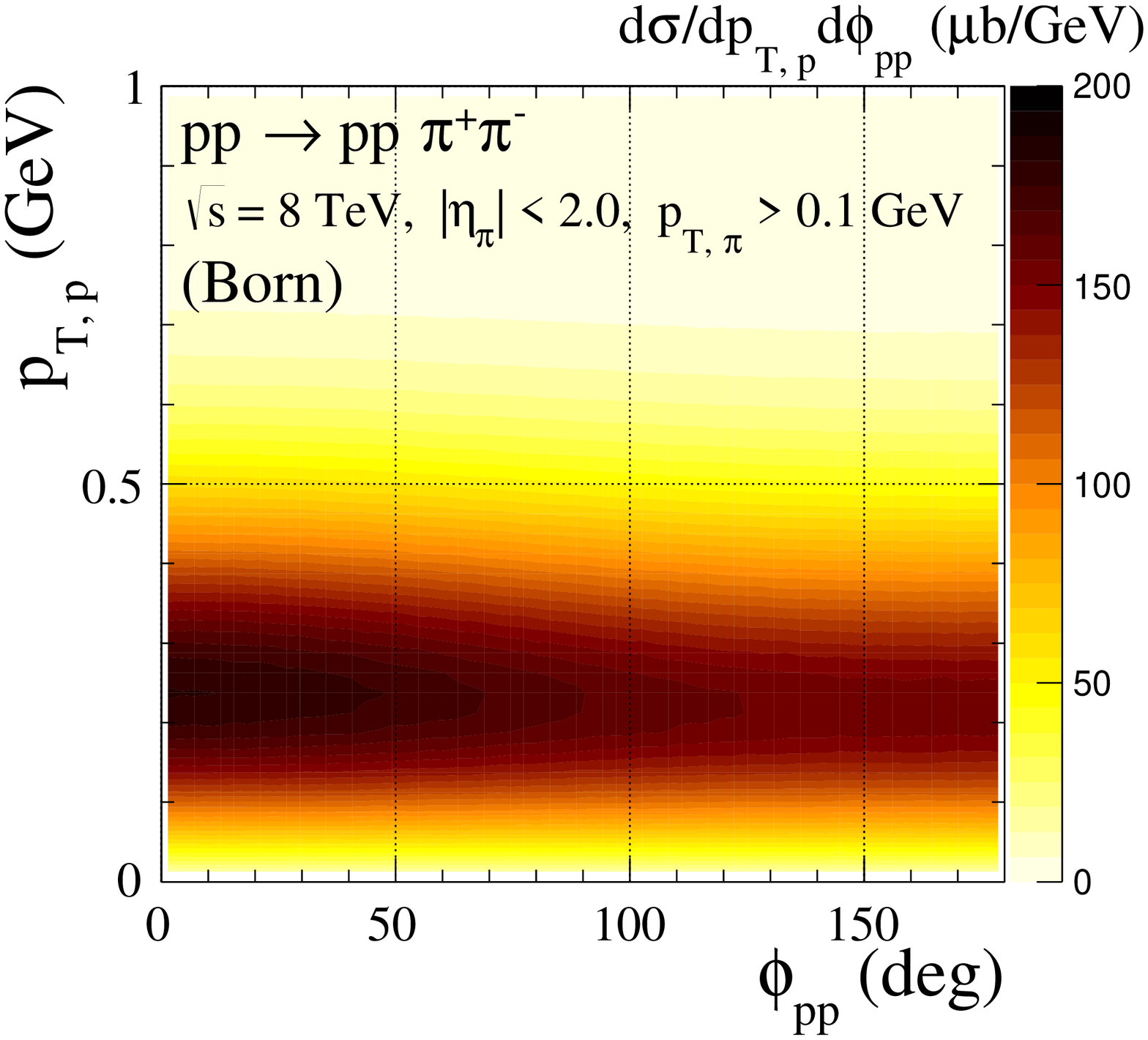}
\includegraphics[width=0.45\textwidth]{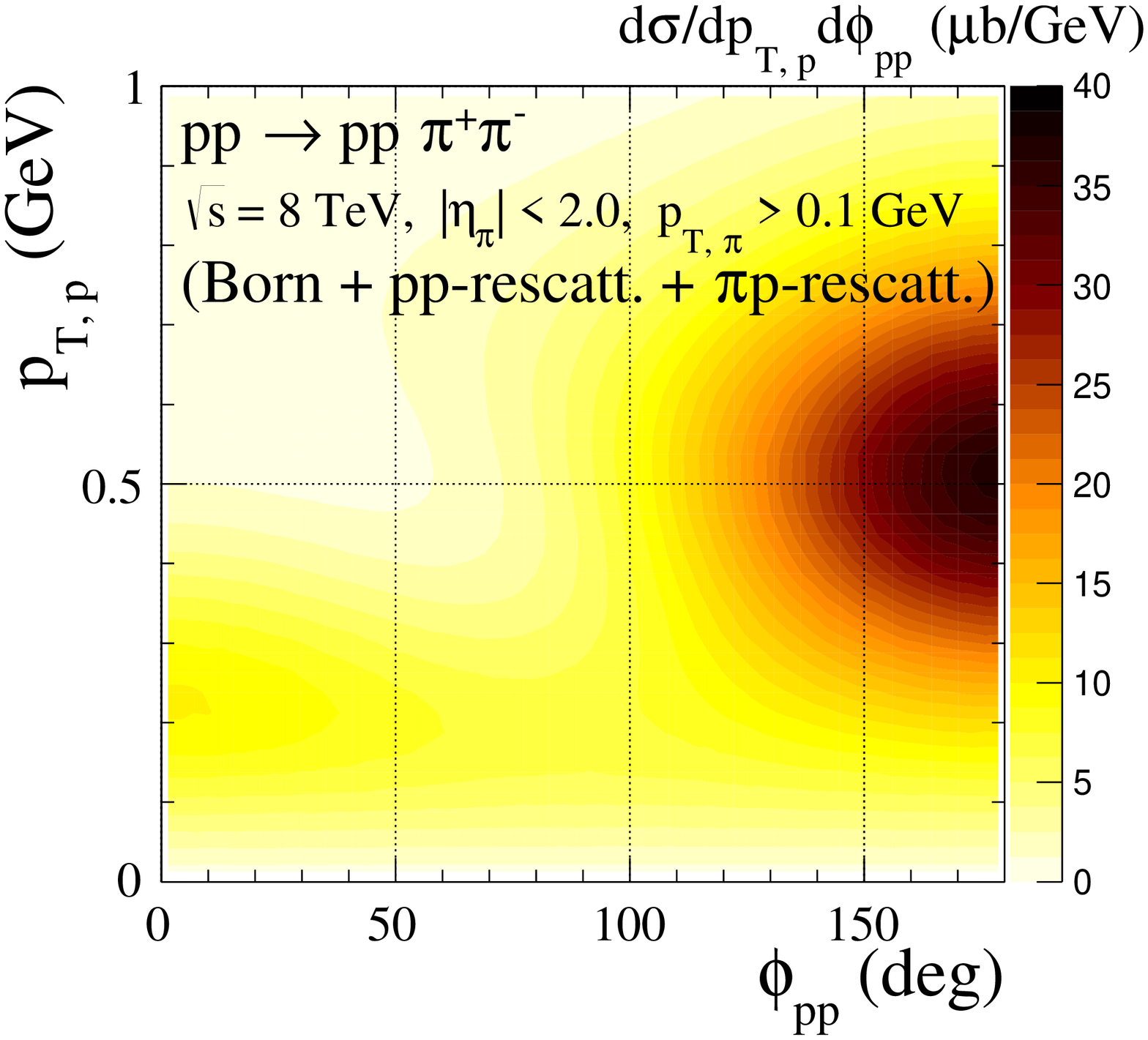}
  \caption{\label{fig:CMS_map_pt1phi12}
  \small
Two dimensional distributions in $p_{t,p}$ and $\phi_{pp}$
at $\sqrt{s} = 8$~TeV with the CMS kinematical cuts specified in the figure caption.
We show the distributions without and with the absorption corrections.
In the calculation we have used the cut-off parameter 
$\Lambda_{off,E} = 1.6$~GeV.
}
\end{figure}

Both the CMS (when combined with TOTEM) and ATLAS (when combined with
ALFA) collaborations can measure outgoing protons.
What additional information can be provided by measuring 
the momenta of the outgoing protons?
In Figs.~\ref{fig:CMS_pt1_phi12} and \ref{fig:CMS_map_pt1phi12} 
we show the influence of the absorption effects on the $p_{t,p}$,
$\phi_{pp}$, and $t$ distributions. 
The distribution in proton transverse momenta are particularly interesting. 
The extra absorption effects due to $\pi p$ interactions make the distributions much broader
than in the case of Born approximation and even broader than
in the case when only $pp$ absorption effects are included.
The effect depends on the value of cut-off parameter $\Lambda_{off}$.
Therefore we expect that the CMS and ATLAS experimental groups could verify our predictions. 
The extra absorption effects lead to significant modification
of the shape of proton-proton relative azimuthal angle distributions
which also could be tested by the two experiments.
\footnote{Note that in the case of the ATLAS experiment the requirement of both protons being
tagged in the ALFA detectors influences the shapes of the distributions only
very little, but it reduces the cross section by a factor close to 3 \cite{Staszewski:2011bg}.}
The distributions in proton four-momentum transfer $t = t_{1} = t_{2}$ 
are presented in Fig.~\ref{fig:CMS_pt1_phi12} (bottom panels).
The extra pion-proton interaction increases the distribution at large $|t|$. 

The effect of absorption can be even better seen in two-dimensional
distributions in proton-proton relative azimuthal angle and
transverse momentum of one of the protons, see \ref{fig:CMS_map_pt1phi12}). 
Quite different pattern can be seen for the Born case and 
for the case with full absorption. 
It is not clear to us whether such a two-dimensional distribution 
can be obtained in practice.

In Table \ref{tab:table} we have collected cross sections in $\mu b$
for the exclusive $\pi^{+}\pi^{-}$ production with absorption effects
discussed in section~\ref{sec:absorption_corrections}
and for some kinematical cuts specified in section~\ref{sec:predictions}.
The Born cross sections for $\sqrt{s}$ = 0.2, 1.96, 7, 8 TeV
and $\Lambda_{off,E}=1.6$~GeV are 1.13, 39.60, 54.71, 192.49 $\mu b$, respectively.
Thus the ratio of full and Born cross sections 
$\langle S^{2}\rangle$ (the gap survival factor)
is approximately 0.20 (STAR), 0.09 (CDF), 0.12 (LHC).
Results at $\sqrt{s} = 13$~TeV were obtained also with the CMS kinematical cuts.
\begin{table}
\begin{footnotesize}
\caption{The integrated cross sections in $\mu b$ 
for the central exclusive $\pi^{+}\pi^{-}$ production
via the double-pomeron/$f_{2 \Reg}$ exchange mechanism
including the $NN$ and $\pi N$ absorption effects.
The results for different experiments with cuts specified in section \ref{sec:predictions}
and for the different values of the off-shell-pion form-factor parameters 
in Eqs.~(\ref{off-shell_form_factors_exp}) and (\ref{off-shell_form_factors_mon}) are shown.
}
\label{tab:table}
\begin{center}
\begin{tabular}{|c|c|c|c|c|c|}
\hline
$\sqrt{s}$~(TeV):          & 0.2    &   1.96    &   7   &   8  &   13    \\
\hline
$\Lambda_{off,E}=1.6$~GeV  & 0.23  & 3.69 & 6.57 & 23.92 & 28.64\\
$\Lambda_{off,E}=1.0$~GeV  & 0.09  & 0.63 & 2.16 &  \;\,7.88 &  \;\,8.98\\
\hline
$\Lambda_{off,M}=1.6$~GeV  & 0.26  & 6.45 & 9.12 & 33.60 & 40.92\\
$\Lambda_{off,M}=0.8$~GeV  & 0.07  & 0.58 & 1.74 &  \;\,6.48 &  \;\,7.45\\
\hline
\end{tabular}
\end{center}
\end{footnotesize}
\end{table}

\section{Conclusions}

In the present paper we have taken into account absorption corrections
due to pion-nucleon final state interaction
in addition to those due to proton-proton interactions. 
To make realistic predictions of the cross sections 
the parameter responsible for off-shellness of intermediate pions 
in the Lebiedowicz-Szczurek model has been adjusted to experimental data. 
We have considered here two different scenarios:
\\
1) The parameters have been adjusted to the STAR data \cite{Adamczyk:2014ofa}, 
where protons have been registered, which guarantees exclusivity of 
the process, however, the statistics was rather low 
and only low dipion invariant masses ($M_{\pi \pi} < 1.5$~GeV) could be observed.
\\
2) The parameters have been adjusted to the CDF data
\cite{Aaltonen:2015uva} (see also \cite{Albrow_Project_new}), 
where only some rapidity gaps outside of the main detector 
was imposed in the experiment.

The cross section for the invariant masses $M_{\pi \pi} < 1$~GeV is subjected to low-energy 
pion-pion final state interaction ($\pi \pi$ FSI) effects 
which are not included in the present analysis.
Thus, in the first scenario, one finds rather large $\Lambda_{off} \approx 1.6$~GeV
in the region of $M_{\pi \pi} < 1$~GeV. 
In the second scenario, when the CDF data are fitted
in the broad range of $M_{\pi \pi}$ one obtains $\Lambda_{off} \approx 0.8$~GeV.
Then as a consequence one underestimates the RHIC data at $M_{\pi \pi} \sim 0.5-1.0$~GeV.
But this missing strength at low $M_{\pi \pi}$ is probably 
due to the $\pi \pi$ FSI enhancement in the $\sigma$ meson region, see \cite{Pumplin:1976dm,Au:1986vs}.
Therefore, we might expected that at higher masses the non-resonant model 
(with no reggeization of intermediate pion) gives realistic predictions
with the off-shell pion parameter $\Lambda_{off} \approx 1.0$~GeV.
We have proposed to use a 'stretched exponential'
parametrization of $\pi N$ amplitudes which better describes the large-$t$ region 
and coincides with the exponential parametrization in small-$t$ region. 
Such a parametrization is more adequate when focussing
on larger transverse momenta.
However, we fail to describe the CDF data with $p_{t,\pi\pi} > 1$~GeV.
Clearly final tuning of the model requires to take into
account both $\pi \pi$ FSI effects as well as explicit resonances
such as the tensor $f_2(1270)$ meson. 
This goes beyond the scope of the present paper, 
where we have concentrated on the absorption effects. 
This will be a subject of our future studies.

However, even the present rather simplified treatment of the reaction
mechanism allows to draw interesting conclusions as far as 
the absorption effects are considered.
The inclusion of the pion-nucleon interactions lead to additional 
damping of the cross section by a factor of about 2,
almost independent of center-of-mass energy at least in the range 
considered in the present paper.
The additional interaction changes the shape of some distributions 
($d \sigma / dt_{1/2}$, $d \sigma / dp_{t,p}$, $d \sigma/d \phi_{pp}$) 
but leaves almost unchanged shape of other distributions 
($d \sigma / dM_{\pi \pi}$, $d \sigma / d y_{\pi}$, 
$d \sigma /d p_{t,\pi}$, $d \sigma/ d \phi_{\pi \pi}$).
Particularly spectacular modifications are obtained for $|t|$ and $p_{t,p}$ distributions.
In particular, a measurement of the distribution 
in the relative azimuthal angle between the $p_{t}$ vectors 
of the outgoing protons can provide a fully differential test 
of the soft survival factors.
This could be verified in future in experiments when both protons 
are measured such as ATLAS + ALFA \cite{Staszewski:2011bg}
or CMS + TOTEM \cite{Osterberg:2014mta}.
In summary, the additional absorption effect discussed here seems 
crucial in detailed understanding of results of ongoing and planned 
experimental investigations. 

\acknowledgments
We are indebted to Mike Albrow, Lidia G\"orlich, Valery Khoze,
Wolfgang Sch\"afer, Reiner Schicker, and Jacek Turnau for interesting discussions.
We are grateful to Maria \.Zurek for sending us the recent CDF data points.
The work of P.L. was supported by the NCN Grant No. DEC-2013/08/T/ST2/00165,
the MNiSW Grant No. IP2014~025173 ``Iuventus Plus''
and by the START fellowship from the Foundation for Polish Science.
The work of A.S. was partially supported 
by the Centre for Innovation and Transfer of Natural Sciences 
and Engineering Knowledge in Rzesz\'ow.

\nocite{}

{
\begin{small}
\bibliography{refs}
\end{small}
}

\end{document}